\documentclass[a4paper,fleqn,usenatbib]{mnras}

\usepackage{newtxtext,newtxmath}
\usepackage[T1]{fontenc}
\usepackage{ae,aecompl}

\usepackage{graphicx}	% Including figure files
\usepackage{amsmath}	% Advanced maths commands
\usepackage{amssymb}	% Extra maths symbols
\usepackage{todonotes}
\usepackage{bm}         % bold maths
\usepackage{comment}    % comment sections
\usepackage{wasysym}     % for solar system symbols
\usepackage{subfig}
\newcommand{\dd}[0]{\mathrm{d}}

\newcommand{\brac}[1]{\left\langle #1 \right\rangle}
\newcommand{\vrac}[1]{\left\vert #1 \right\vert}
\newcommand{\wrac}[1]{\left\| #1 \right\|}
\newcommand{\rH}[0]{r_{\mathrm{H}}}
\newcommand{\rB}[0]{r_{\mathrm{B}}}

\newcommand{\vK}[0]{v_{\mathrm{K}}}
\newcommand{\low}[1]{_{\mathrm{#1}}}

\def\ba{\begin{eqnarray}}
\def\ea{\end{eqnarray}}

\graphicspath{ {./} }

\title[Envelopes of embedded super-Earths II]{Envelopes of embedded super-Earths\\II. Three-dimensional isothermal simulations}

\author[W. B\'ethune \& R. R. Rafikov]{
William B\'ethune$^{1}$\thanks{E-mail: wb288@damtp.cam.ac.uk},
Roman Rafikov$^{1,2}$
\\
$~^{1}$ Department of Applied Mathematics and Theoretical Physics, University of Cambridge, \\
Centre for Mathematical Sciences, Wilberforce Road, Cambridge CB3 0WA, UK.\\
$~^{2}$ Institute for Advanced Study, Einstein Drive, Princeton, NJ 08540
}

\date{Accepted XXX. Received YYY; in original form ZZZ}

\pubyear{2018}

\begin{document}
\label{firstpage}
\pagerange{\pageref{firstpage}--\pageref{lastpage}}
\maketitle

\begin{abstract}
Massive planetary cores embedded in protoplanetary discs are believed to accrete extended atmospheres, providing a pathway to forming gas giants and gas-rich super-Earths. The properties of these atmospheres strongly depend on the nature of the coupling between the atmosphere and the surrounding disc. We examine the formation of gaseous envelopes around massive planetary cores via three-dimensional inviscid and isothermal hydrodynamic simulations. We focus the changes in the envelope properties as the core mass varies from low (sub-thermal) to high (super-thermal) values, a regime relevant to close-in super-Earths. We show that global envelope properties such as the amount of rotational support or turbulent mixing are mostly sensitive to the ratio of the Bondi radius of the core to its physical size. High-mass cores are fed by supersonic inflows arriving along the polar axis and shocking on the densest parts of the envelope, driving turbulence and mass accretion. Gas flows out of the core's Hill sphere in the equatorial plane, describing a global mass circulation through the envelope. The shell of shocked gas atop the core surface delimits regions of slow (inside) and fast (outside) material recycling by gas from the surrounding disc. While recycling hinders the runaway growth towards gas giants, the inner regions of protoplanetary atmospheres, more immune to mixing, may remain bound to the planet. 
\end{abstract}


%%%%%%%%%%%%%%%%%%%%%%%%%%%%%%%%%%%%%%%%%%%%%%%%%%

%%%%%%%%%%%%%%%%% BODY OF PAPER %%%%%%%%%%%%%%%%%%

\section{Introduction}

Over $4000$ planetary candidates have been discovered by the Kepler mission \citep{borucki10,batalha13}. Simultaneous measurements of masses and radii available for some planets give insights about their composition, and reveal a population of `super-Earths' --- planets with a mass in the range $(1-10)~M_\oplus$ and a radius in the range $(1-4)~R_\oplus$. Given the low average density of many such planets \citep{weissmarcy14,rogers15}, up to tens of percent of the mass of super-Earths could reside in their gaseous atmospheres \citep{lopezfortney14,wolfganglopez15}.

In one-dimensional models of embedded planetary envelopes, a planet can accrete as much gas as radiative cooling permits \citep{leechiang15}. Eventually, no static equilibrium can be maintained beyond a critical core mass $\sim 5 - 20~M_{\oplus}$ \citep{perricameron74,mizuno78,rafikov06}, leading to a phase of runaway gas accretion. This scenario is successful in forming gas giant planets \citep{mizuno80,pollack96}, but it also suggests that, unless forming late in the life of the disc \citep{leechiang14,leechiang16}, many super-Earths should have turned into gas giants over the lifetime of the disc. 

One way to counteract the radiative cooling of planetary envelopes and to prevent runaway core accretion is to continuously replace their material by high-entropy gas from the disc \citep{ormel2}. The efficiency of this `recycling' process motivated dedicated studies of the flow structure around planetary cores \citep{fung15,kurokawa18}, which confirm the potential importance of this mechanism. In \citet[][hereafter Paper I]{bethunerafi19a} we explored the properties of embedded planetary envelopes using two-dimensional (2D) simulations. In this second paper of a series, we extend the 2D results of Paper I to three dimensions by running a suite of 3D simulations of the flow near the core. We explore how the flow properties change as the core mass varies across the so-called thermal mass scale \citep{rafikov06}. We restrict ourselves to isothermal simulations and postpone a more realistic treatment of the thermodynamics to a later study. 

As part of describing 3D envelopes,  we will primarily focus on differences in flow morphology, mass accretion, the formation of rotationally supported envelopes, and the efficiency of gas recycling as a function of core mass. We will also systematically explore the differences with the 2D results brought about by extending the flow in the vertical dimension. Compared to \cite{bate03}, we will focus on scales smaller than the pressure scale height of the disc, and will consider gas accretion in an inviscid context. 

Another subject of our study is the role of the finite size of the core in determining the flow properties around the core. To handle the large contrast between the core radius and the stratification scale of the disc, many previous studies used a softened gravitational potential and/or sink cells in place of the core \citep[e.g.,][]{kley01,bate03,machida10,tanigawa12}. \cite{ormel1} and \cite{fung14} noticed the importance of properly resolving the core scale, and its influence on the entire envelope, as we verified in 2D in Paper I. We follow the same approach here and fully account for the boundary condition imposed by the core surface. 

The paper is organised as follows. We present our physical setup and numerical model in \autoref{sec:method}. The results are described starting with a reference case in \autoref{sec:results} and then sampling different values for the core mass and size in \autoref{sect:res_vary}. We discuss these results focusing on the differences between the 2D and 3D situations in \autoref{sec:discussion}. Our main findings are summarised in \autoref{sec:summary}. 

%%%%%%%%%%%%%%%%%%%%%%%%%%%%%%%%%%%%%%%%%%%%%%%%%%
%%%%%%%%%%%%%%%%%%%%%%%%%%%%%%%%%%%%%%%%%%%%%%%%%%

\section{Method}
\label{sec:method}

%%%%%%%%%%%%%%%%%%%%%%%%%%%%%%%%%%%%%%%%%%%%%%%%%%

The method employed to address the current problem is similar to that in Paper I, with the obvious distinction that the flow can now evolve in three dimensions. We will therefore focus on the main distinctions from the 2D case; Paper I should be consulted for more details.

%%%%%%%%%%%%%%%%%%%%%%%%%%%%%%%%%%%%%%%%%%%%%%%%%%

\subsection{Physical model}

%%%%%%%%%%%%%%%%%%%%%%%%%%%%%%%%%%%%%%%%%%%%%%%%%%

\subsubsection{Notations and main equations}

We model the 3D flow of the disc fluid in the vicinity of a planetary core with mass $m_c$ and radius $r_c$ orbiting the central star $m_\star$ on a circular orbit with a semi-major axis $a$, so that the corresponding angular frequency is $\Omega=(Gm_\star/a^3)^{1/2}$. Let $\rho$ be the gas density, $P$ its pressure and $v$ its velocity. We adopt an isothermal equation of state $P(\rho)=\rho c_s^2$ with a constant sound speed $c_s$. 

We study the dynamics of the disc fluid in a rotating frame centered on the planetary core in the `local' approximation of a Keplerian disc \citep{hill78}. Let $(x,y,z)$ be Cartesian coordinates measured from the center of the core, with $x$ along the radius from the star to the core, $y$ along the orbit of the core and $z$ normal to the disc midplane. Let $(r,\theta,\varphi)$ be spherical coordinates centered on the core, with the polar axis $\theta=0$ along the $z$ direction and $\varphi=0$ along the $x$ direction. We also make use of cylindrical coordinates $(R,\varphi,z)$, with $R=r \sin\theta$ being the cylindrical radius measured from the polar ($z$) axis. 

The gas density and momentum evolve according to
\begin{alignat}{3}
  &\partial_t \rho &&+ \nabla\cdot\left[\rho \bm{v}\right] &&= 0\,,\label{eqn:consrho}\\
  &\partial_t \left[\rho \bm{v}\right] &&+ \nabla\cdot\left[ \rho \bm{v\otimes v} + P \,\mathbb{I} \right] &&= - \rho \nabla \Phi - 2 \rho \bm{\Omega\times v}\,, \label{eqn:consrhov}
\end{alignat}
where $\Phi$ is the full gravitational potential detailed below.

%%%%%%%%%%%%%%%%%%%%%%%%%%%%%%%%%%%%%%%%%%%%%%%%%%

\subsubsection{Gravitational potential}

The gravitational potential of the star is expanded to second order about the orbital radius of the core. It is also expanded to second order in the vertical direction, so the disc will be stratified on a pressure scale $h$. The total gravitational potential $\Phi$ is 
\begin{equation} 
\label{eqn:totpotgrav}
  \Phi = -\frac{G m_c}{r} + q\Omega^2 x^2 + \frac{1}{2}\Omega^2 z^2,
\end{equation}
where we assume a Keplerian shear $q=-3/2$. Unlike Paper I, in this work we use the exact Newtonian potential without any softening near the core surface. The transition from the star's tidal potential to the core-dominated potential occurs at one Hill radius 
\begin{equation}
\rH \equiv \left(\frac{G m_c}{3 \Omega^2}\right)^{1/3}=a\left(\frac{m_c}{3m_\star}\right)^{1/3}
\label{eq:rH}
\end{equation}
away from the core in the equatorial plane. Rotational support of the gas in the envelope around the core will be measured via the Keplerian acceleration $a\low{K} \equiv G m_c/r^2$ and the Keplerian velocity $\vK\equiv \sqrt{G m_c/r}$ due to the core only. 

In the absence of the core (i.e. with $m_c=0$) the potential in the form \eqref{eqn:totpotgrav} allows for a steady shear flow in the disc with 
\ba   
v_x=v_z=0,~~v_y = q\Omega x,~~\rho = \rho_0 \exp\left(-z^2 / 2 h^2\right),
\label{eq:shear_flow}
\ea  
assuming that the midplane density $\rho_0$ is constant in space. As in Paper I, we neglect the headwind arising in sub-Keplerian discs and caused by global pressure gradients; its effect has been considered by, e.g., \citet{ormel2,kurokawa18}.

%%%%%%%%%%%%%%%%%%%%%%%%%%%%%%%%%%%%%%%%%%%%%%%%%%

\subsubsection{Characteristic space and time scales}

The sound speed sets the characteristic pressure scale height of the disc $h\equiv c_s / \Omega$ as well as the Bondi radius of the core $\rB \equiv G m_c / c_s^2$. In the absence of other characteristic length scales, the properties of the flow around a planetary core are expected to be determined by $\rB/h$. This ratio coincides with the ratio of the core mass to its `thermal mass' $m_{\rm th}$ and is closely related to the Hill radius \citep{rafikov06}: 
\ba   
\frac{\rB}{h}=\frac{m_c}{m_{\rm th}}=3\left(\frac{\rH}{h}\right)^3,
\label{eq:mth-rel}
\ea  
see definition \eqref{eq:rH}.

The physical size of the core $r_c$ is an additional free length scale of the problem, which we will always assume being small compared to the pressure scale, $r_c\ll h$. In our simulations the core is represented by a spherical boundary at $r=r_c$, with the requirement that matter should not flow through it, see \autoref{sec:initboundcond}. The core is capable of inducing substantial perturbation in the disc flow via its gravity whenever $\rB\gtrsim r_c$, which sets a lower limit on the core mass of 
\ba   
&& m_c>m_{\rm B} \equiv  m_\star\left(\frac{h}{a}\right)^3\left(\frac{3}{4\pi}\frac{m_\star}{\rho_c a^3}\right)^{1/2}
\label{eq:mB}\\
%%%%%%%%%
&& = 0.007m_\oplus~\left(\frac{m_\star}{m_\odot}\right)^{3/2}
\left(\frac{a}{0.1\mbox{AU}}\right)^{-1}
\left(\frac{h/a}{5\%}\right)^{3}
\left(\frac{\rho_c}{5~\mbox{g cm}^{-3}}\right)^{-1/2},
\nonumber
\ea  
where $\rho_c$ is the bulk density of the core. In this work we will always be in the regime $m_c > m_{\rm B}$, focusing primarily on $m_c\sim m_{\rm th}$

As in Paper I, we find it convenient to introduce dimensionless length scales by normalizing $\rB$ and $h$ by $r_c$:
\ba
B\equiv \frac{\rB}{r_c},~~~~~~H\equiv\frac{h}{r_c}. 
\label{eq:dim-less}
\ea
By our assumptions $B>1$ (as $m_c>m_{\rm B}$) and $H\gg 1$. Numerical estimates of these parameters typical for \emph{Kepler} super-Earths can be found in Paper I. 

The time required to restore a quasi-static equilibrium through the envelope is the sound crossing-time across one pressure scale $\sim \Omega^{-1}$ \citep{miki82}. However, deep near the core  hydrodynamic fluctuations evolve on time scales $\sim r_c/c_s = (\Omega H)^{-1}$. This scale separation makes it computationally expensive to integrate \eqref{eqn:consrho} and \eqref{eqn:consrhov} over more than a few tens of orbital times ($2\pi/\Omega$) of the core around the central star. Such timescales are short relative to planet migration or disc clearing, so our results should be interpreted as a quasi-instantaneous snapshot in the planet formation history.

%%%%%%%%%%%%%%%%%%%%%%%%%%%%%%%%%%%%%%%%%%%%%%%%%%

\subsection{Numerical method} 
\label{sec:numscheme}

%%%%%%%%%%%%%%%%%%%%%%%%%%%%%%%%%%%%%%%%%%%%%%%%%%

Our numerical approach is to evolve equations \eqref{eqn:consrho} and \eqref{eqn:consrhov} in spherical $(r,\theta,\varphi)$ geometry, naturally extending the calculations done in Paper I to a third dimension. Modifications introduced by this extension are discussed next.

%%%%%%%%%%%%%%%%%%%%%%%%%%%%%%%%%%%%%%%%%%%%%%%%%%

\subsubsection{Integration scheme}

The numerical method is largely the same as in Paper I. We use the finite-volume code PLUTO 4.0 \citep{mignone07} to integrate \eqref{eqn:consrho} and \eqref{eqn:consrhov} in conservative form. Primitive variables are estimated at cell interfaces by linear reconstruction with VanLeer's slope limiter \citep{vanleer79}. Godunov fluxes are computed via the Roe approximate Riemann solver \citep{roe81}. The equations are integrated in time via an explicit second order Runge-Kutta scheme; the Courant-Friedrichs-Lewy (CFL) stability criterion is satisfied by timesteps $\Delta t \leq 0.3 \Delta x/ U$ for characteristic velocities $U$ over a grid spacing $\Delta x$. The Coriolis acceleration is included in a conservative fashion by integrating \eqref{eqn:consrhov} in a frame rotating with angular frequency $\Omega$ about the polar axis \citep{kley98,mignone12}. 
%%%%%%%%%%%%%%%%%%%%%%%%%%%%%%%%%%%%%%%%%%%%%%%%%%

\subsubsection{Computational domain}

The computational domain is defined in spherical geometry by $\left(r,\theta,\varphi \right) \in \left[r_c,128 r_c\right] \times \left[0,\pi\right] \times \left[0,2\pi\right]$. It is meshed by $128 \times 80 \times 160$ grid cells with a logarithmic spacing in the radial direction and constant spacings in the other dimensions. This moderate resolution will allow us to perform a series of simulation runs at a reasonable computational cost. The dependence of most diagnostics on resolution was examined in Appendix A of Paper I, where convergence was found starting at $128$ cells in the radial direction. This represents approximately $18$ cells between $[r,2r]$ for any $r$. 
%%%%%%%%%%%%%%%%%%%%%%%%%%%%%%%%%%%%%%%%%%%%%%%%%%

\subsubsection{Stabilizing procedures}

The numerical scheme does not intrinsically preserve the positivity of density, so we provide additional dissipation to stabilize the solver in extreme conditions. If the relative pressure variations exceed a factor of $10$ between neighboring cells, or if the local Mach number $v/c_s \geq 10$, we switch to the MINMOD slope limiter and to the HLL approximate Riemann solver in the concerned cells \citep{van1997relation}.

For simulations with $H=16$, the computational domain goes as high as eight pressure scale heights above the midplane. In hydrostatic equilibrium, the density given by (\ref{eq:shear_flow}) becomes extremely low at such altitudes. Small momentum fluctuations easily result in high velocities, turbulence and shocks in the upper layers of the disc. Because our interest is on the dense parts of the disc, the gravitational potential of the star $\Omega^2 z^2 / 2$ is set to a constant value $8\Omega^2 h^2$ for $z>4h$. In the absence of a core, the hydrostatic disc density would therefore be constant above four pressure scale heights. 

%%%%%%%%%%%%%%%%%%%%%%%%%%%%%%%%%%%%%%%%%%%%%%%%%%

\subsubsection{Initial and boundary conditions} 
\label{sec:initboundcond}

The initial conditions consist of the background, unperturbed shear flow of the disc in the form \eqref{eq:shear_flow}, where we additionally extrapolate $\rho$ to a constant for $z\geq 4h$. To prevent a violent relaxation of the flow toward the core starting from this initial state, the core mass $m_c$ is linearly increased from zero to its nominal value over the first $\Omega t / 2\pi = 2$ orbital times. As soon as the core potential is fully introduced, the properties of the envelope evolve on a much longer timescale (see Figure A.1 of Paper I). 

The azimuthal boundary conditions are periodic. The conditions in the polar dimension respect the spherical topology: about the axis $\theta=0\pm\vartheta$ in one hemisphere, 
\begin{equation}
\left[\rho,v_r,v_{\theta},v_{\varphi}\right](r,-\vartheta,\varphi) = \left[+\rho,+v_r,-v_{\theta},-v_{\varphi}\right](r,\vartheta,\varphi+\pi), 
\end{equation}
and similarly in the opposite hemisphere about $\theta=\pi\pm\vartheta$. 

At the outer radial boundary, we prescribe the initial, unperturbed state in the ghost cells. Because this condition does not smoothly match the conditions inside the computational domain, a discontinuity appears in the outermost active grid cells. We find no noticeable effect of this discontinuity on the flow near the core; to avoid possible effects of the outer boundary on our results, the outer $R>64r_c$ will always be excluded from our analysis. We emphasize that disc-scale processes such as the formation of a density gap cannot be captured in our model without ad-hoc prescriptions at the outer radial boundary. 

At the inner radial boundary, we prevent mass inflows by the adequate symmetrization in the ghost cells
\begin{equation}
\left[\rho,v_r,\Phi_c\right](r_c-x) = \left[+\rho,-v_r,+\Phi_c\right](r_c+x). 
\end{equation}
Enforcing an even symmetry for $\Phi_c$ helps cancelling the gravitational acceleration at the interface. We monitor the mass losses through this boundary by integrating cell-averaged mass fluxes. The mass losses through the inner radial boundary are negligible to better than $10^{-2}$ accuracy compared to the mass accretion rates through the envelope, so the core surface is indeed impermeable. 

The choice of $\left(v_{\theta},v_{\varphi}\right)$ in the inner radial ghost cells will likely influence the long-term evolution of the envelope. In the absence of explicit viscosity, the spin of the core can only affect its envelope via turbulent or numerical diffusion. In Paper I, we used a two-dimensional conservation argument to prescribe $v_{\varphi}$ in a stress-free fashion. There is no simple and equivalent prescription in three dimensions. We opt for simplicity to make boundary effects easier to identify. The meridional velocity is set to $v_{\theta}=0$ in the inner radial ghost cells. The azimuthal velocity $v_{\varphi}$ could be influenced by a number of torques resulting, for example, from the angular momentum of accreted planetesimals or from tidal synchronization with the star. We set $v_{\varphi} = (q/2)\, r \sin\left(\theta\right)$, i.e. a solid-body rotation with the same vorticity as in the Keplerian flow. This corresponds to the steady state of a solid sphere under the friction of a viscous shear flow. 
%%%%%%%%%%%%%%%%%%%%%%%%%%%%%%%%%%%%%%%%%%%%%%%%%%

\subsection{Units and conventions} 
\label{sec:units}

The orbital frequency $\Omega$ and the core radius $r_c$ are taken as frequency and distance units. The isothermal sound speed of the disc is $c_s = \Omega h = H$ and we take the gravitational constant $G=1$, so the core mass becomes $m_c = B H^2$ in these units. We take the midplane density of the background disc $\rho_0$ as density unit. We label each simulation run by its isothermal pressure scale height \verb|H|\# and its Bondi radius \verb|B|\#. 

In our analysis we will sometimes separate a given quantity $X$ into a `secular' (moving-average) component $\overline{X}$ and a fluctuating component $X^\prime$. The secular term $\overline{X}$ is estimated by averaging over a set of simulation snapshots; it is assumed to vary only moderately on the time interval considered. By default, we use six snapshots spanning three orbits of the core when doing this averaging. The fluctuations are computed by subtracting the average $\overline{X}$ in each snapshot. Spatial averages are denoted with brackets $\brac{\cdot}$. Depending on the context, they will refer to azimuthal averages or to averages on the sphere $\brac{X}_{\mathcal{S}} = \oint_{\mathcal{S}} X \,\dd S / 4\pi r^2$. 

The parameters and main diagnostics of the simulations presented in this paper are listed in \autoref{tab:recap}. The highest $H=32$ considered in this study is still smaller by about a factor of 2 compared to what is expected for super-Earths at 0.1 AU from the star \citep{rogers15}. This restriction comes from the CFL constraint on sound waves at the grid scale near the surface of the core. The confirmation of our findings at larger scale separations $H>32$ and longer timescales $\Omega t / 2\pi > 100$ will require additional computational resources.

We refer to the runs using their values of $H$ and $B$. We take \texttt{H16B16} as our reference (fiducial) case, against which the other simulations will be compared. 

\begin{table}
\caption{Simulations presented in this paper: label, pressure scale height $H$, Bondi radius $B$ (both normalised by $r_c$), maximal percentage of rotational support $v_{\varphi}/\vK$ reached in the equatorial plane, maximal Mach number $v_z/c_s$ reached by the time-averaged flow in a cone of half-opening angle $\pi/16$ about the polar axis.}
\begin{tabular}{lcccc}
\label{tab:recap}
Label & $H$ & $B$ & $\%$ equatorial $v_{\varphi}/\vK$ & polar $v_z/c_s$\\
\hline
\verb!H16B8!  & $16$ & $8$  & $1.20\pm 0.01$ & $0.14 \pm 0.01$ \\
\verb!H16B16! & $16$ & $16$ & $38\pm 1~~$ & $2.4 \pm 0.1$ \\    %max at 3.05r_c, drop through  2.6r_c
\verb!H16B32! & $16$ & $32$ & $86\pm 2~~$ & $5.8 \pm 0.1$ \\    %max at 1.55r_c, drop through 1.45r_c
\verb!H16B64! & $16$ & $64$ & $105\pm 1~~$ & $9.4 \pm 0.1$ \\   %max at  1.3r_c, drop through 1.25r_c
\verb!H32B16! & $32$ & $16$ & $31.3\pm 0.3~$ & $1.20 \pm 0.08$ \\  %max at  3.9r_c, drop through  2.6r_c
\verb!H32B32! & $32$ & $32$ & $85.4\pm 0.6$ & $5.92 \pm 0.03 $ \\ %max at 1.67r_c, drop through  1.6r_c
\end{tabular}
\end{table}

%%%%%%%%%%%%%%%%%%%%%%%%%%%%%%%%%%%%%%%%%%%%%%%%%%
%%%%%%%%%%%%%%%%%%%%%%%%%%%%%%%%%%%%%%%%%%%%%%%%%%

\section{Results: fiducial simulation}
\label{sec:results}

We start by presenting the results of the fiducial simulation run \texttt{H16B16}. As it has $B/H=1$, the core mass equals the thermal mass $m_c=m_{\rm th}$ \citep{rafikov06}, corresponding to the transition from low to high-mass cores. In this regime, gravitational torques caused by the core are expected to trigger a non-linear response in the disc close to the planet \citep{KP96}, i.e. density perturbations at $r\sim h$ away from the planet should be of order $\rho_0$. 

%%%%%%%%%%%%%%%%%%%%%%%%%%%%%%%%%%%%%%%%%%%%%%%%%%

\subsection{Flow structure}
\label{sect:flow_fid}

%%%%%%%%%%%%%%%%%%%%%%%%%%%%%%%%%%%%%%%%%%%%%%%%%%

\begin{figure}%[H]
\begin{center}
\includegraphics[width=1.0\columnwidth]{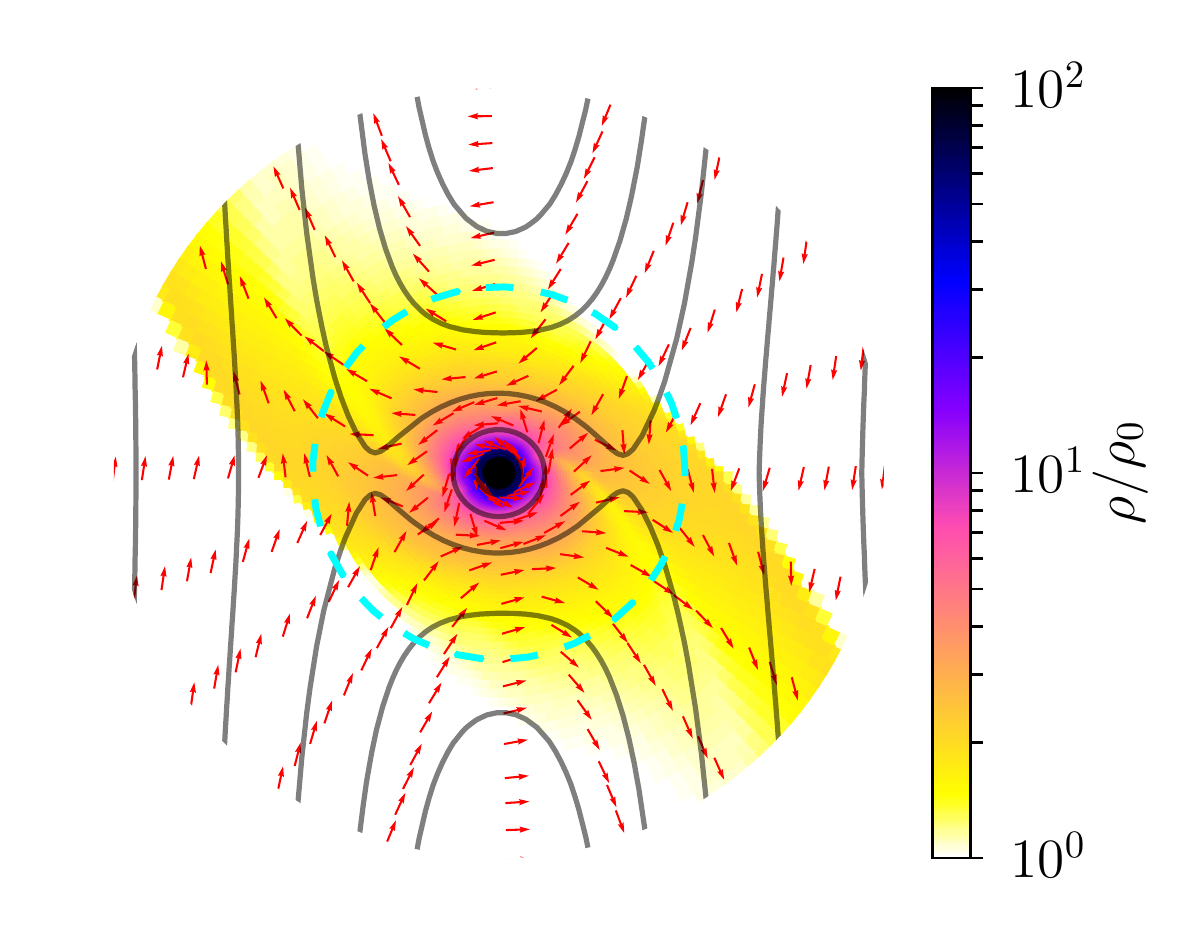}
\caption{Time-averaged density (color map) and velocity (red arrows, orientation only) in the equatorial plane of the fiducial run \texttt{H16B16} with $m_c=m_{\rm th}$, $B=H=16$. The dashed cyan circle marks one pressure scale $h$ centered on the core. Iso-contours of the total gravitational potential $\Phi$ are traced in grey. 
\label{fig:i3dp16b16_EQt_tot}}
\end{center}
\end{figure}

\begin{figure}%[H]
\begin{center}
\includegraphics[width=1.0\columnwidth]{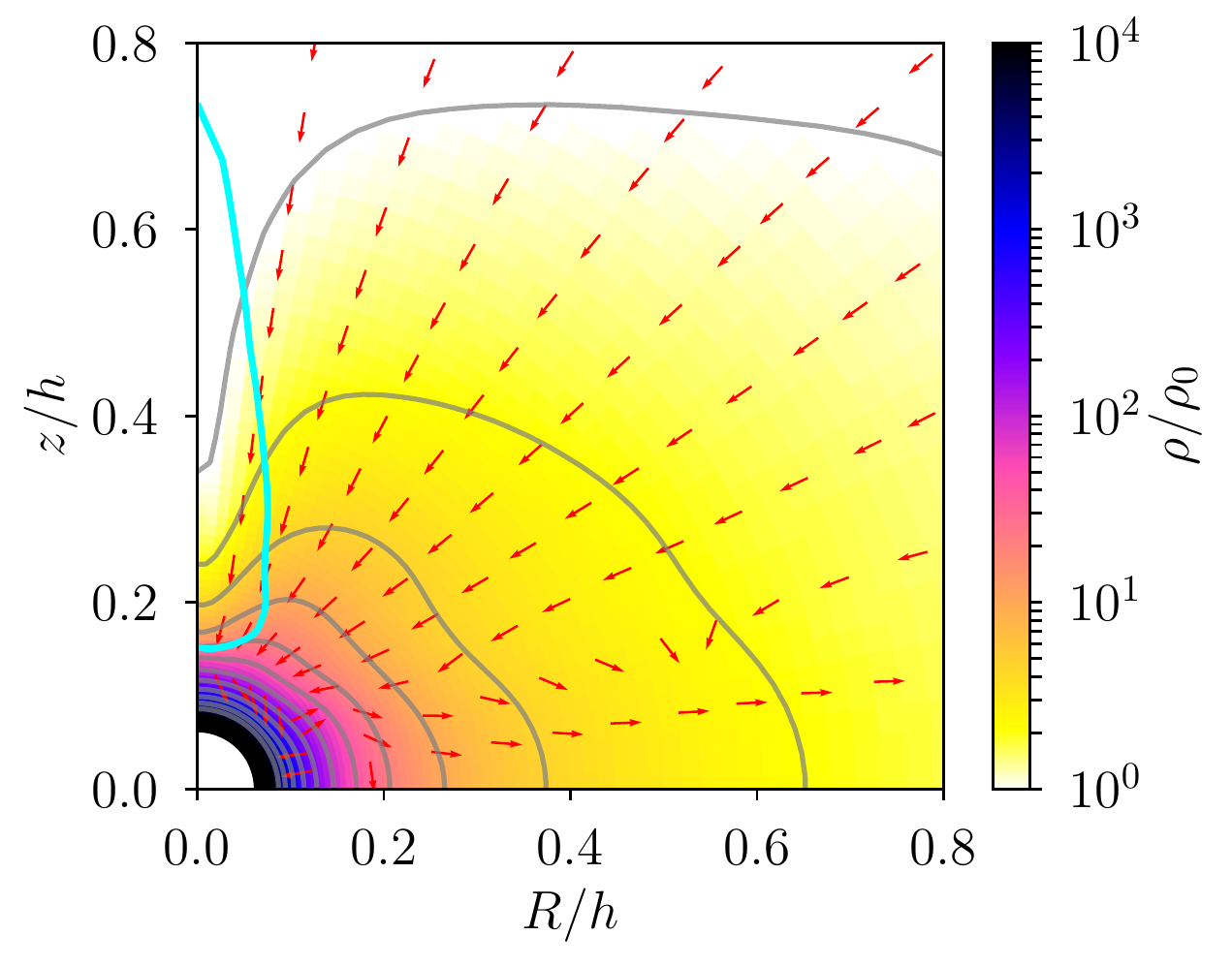}
\caption{Time and azimuthally-averaged density $\rho$ (color map and thin grey contours) and mass flux $\rho v$ (red arrows, orientation only) in the poloidal plane of run \texttt{H16B16}. The thick cyan line near the polar axis marks the sonic surface of the poloidal velocity: the polar inflow smoothly becomes supersonic and shocks close to the core. Only the upper hemisphere is represented here, a zoomed-out view is shown on \autoref{fig:PLm_rhodv}b. \label{fig:i3dp16b16_PLm_rhodv}}
\end{center}
\end{figure}

We start by discussing the time-averaged characteristics of the flow in run \texttt{H16B16}. \autoref{fig:i3dp16b16_EQt_tot} shows the density and velocity distributions in the equatorial plane after averaging over three orbital periods. The flow can be naturally separated into distinct regions \citep{fung15}: the background shear flow (left and right sides), the co-orbital flow completing U-turns (top and bottom), and circular streamlines near the core. The flow in these different regions roughly follows the contours of constant gravitational potential. Density wakes are launched from about one pressure scale away from the core, extending into spiral density waves on larger scales.

In comparison with the 2D case, the gas density ratio between the atmosphere on top of the core and the background disc is several orders of magnitude larger in 3D. This difference is discussed in more detail in \autoref{sect:rot_support}. The closed streamlines orbit the core with a prograde orientation ($v_{\varphi}>0$), but they span a smaller area than the equivalent 2D flow. 

\autoref{fig:i3dp16b16_PLm_rhodv} shows the time-averaged flow in the poloidal plane $(R,z)$ after averaging in the azimuthal direction $\varphi$. Deviations from spherical symmetry clearly appear in the density contours. A key feature obvious in this figure is that the mass flow is oriented toward the core at high latitudes and away from the core in the equatorial plane. Every simulation presented in this paper displays such a global circulation, with the same orientation --- outflow near the equatorial plane --- as found by, e.g, \cite{bate03}, \cite{fung15}, \cite{lamblega17}, \cite{Kuwahara}. The isodensity contours are pinched toward the core near the poles and outward in the equatorial plane. The polar inflow reaches a Mach number $v_z/c_s \approx 2.4$ on the polar axis before shocking on the dense inner envelope, at $z\approx 2.4 r_c$. This transition from supersonic to subsonic poloidal flow happens at the lower part of the sonic surface on \autoref{fig:i3dp16b16_PLm_rhodv}. The maximum Mach number reached in the time-averaged polar inflow of each run is indicated in \autoref{tab:recap}.

%%%%%%%%%%%%%%%%%%%%%%%%%%%%%%%%%%%%%%%%%%%%%%%%%%

\subsection{Quasi-steady state} 
\label{sec:fidqsstate}

At the beginning of a simulation, every diagnostic smoothly evolves as the mass of the core is progressively increased. Once the gravitational potential of the core is fully set, the evolution becomes much slower. Momentum balance is established on sound crossing timescale $h/c_s=\Omega^{-1}$, so a quasi-steady state is effectively achieved after four orbits. Even though the mass of the envelope keeps slowly evolving in time, the mass accretion rate $\dot M$ is roughly constant over the limited integration time of each simulation (after the initial build-up phase during which the core mass is increased).

\begin{figure}%[H]
\begin{center}
\includegraphics[width=1.0\columnwidth]{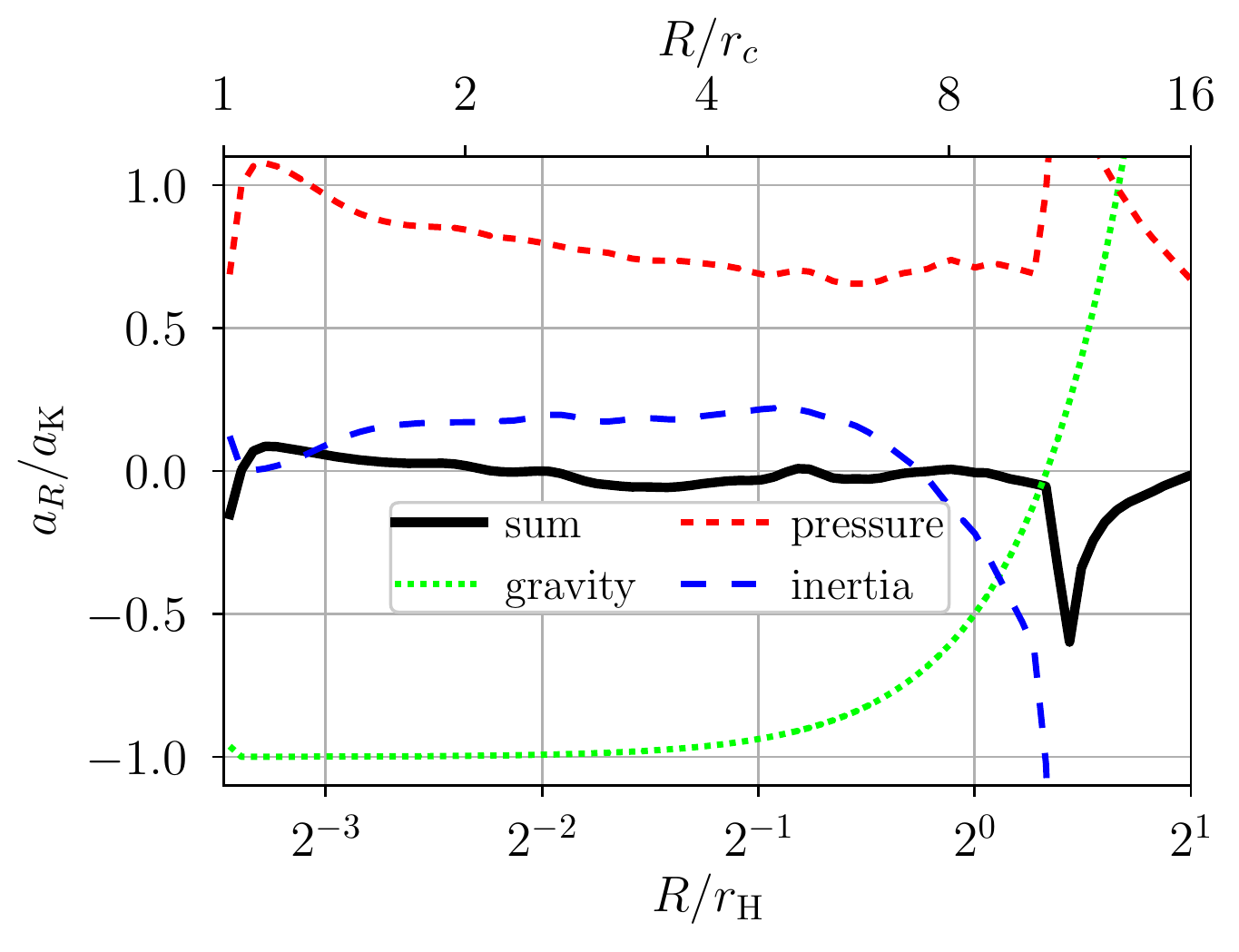}
\caption{Contributions to the radial acceleration relative to Keplerian after time and azimuthal averaging in the equatorial plane of run \texttt{H16B16}; the total acceleration (solid black) is decomposed into gravitational (dotted green), pressure (dashed red) and inertial terms (dashed blue).\label{fig:i3dp16b16_EQ_axelpar}}
\end{center}
\end{figure}

On \autoref{fig:i3dp16b16_EQ_axelpar}, the radial acceleration $a_R\equiv \partial_t v_R$ is compared to the Keplerian acceleration $a\low{K}\equiv m_c / R^2$. We decompose $a_R$ into several components arising in the equation of motion: a gravitational term $-\nabla \Phi$ including the tidal potential of the star, a pressure term $-\nabla P$, and the inertial term including $(v\cdot\nabla ) v$ and the Coriolis acceleration. Each term is evaluated from the time-averaged flow variables and then azimuthally averaged in the equatorial plane. 

The sum of these different terms is close to zero: the flow maintains a radial momentum balance dominated by pressure against gravity. The pressure jump marks the boundary of the envelope at one pressure scale away from the core. An important feature of the momentum balance in 3D is that the rotational support is weaker than in the equivalent 2D case (cf. Figure 3 of Paper I). As a result, pressure support dominates over rotational support in the equatorial plane of this 3D run. 

\begin{figure}%[H]
\begin{center}
\includegraphics[width=1.0\columnwidth]{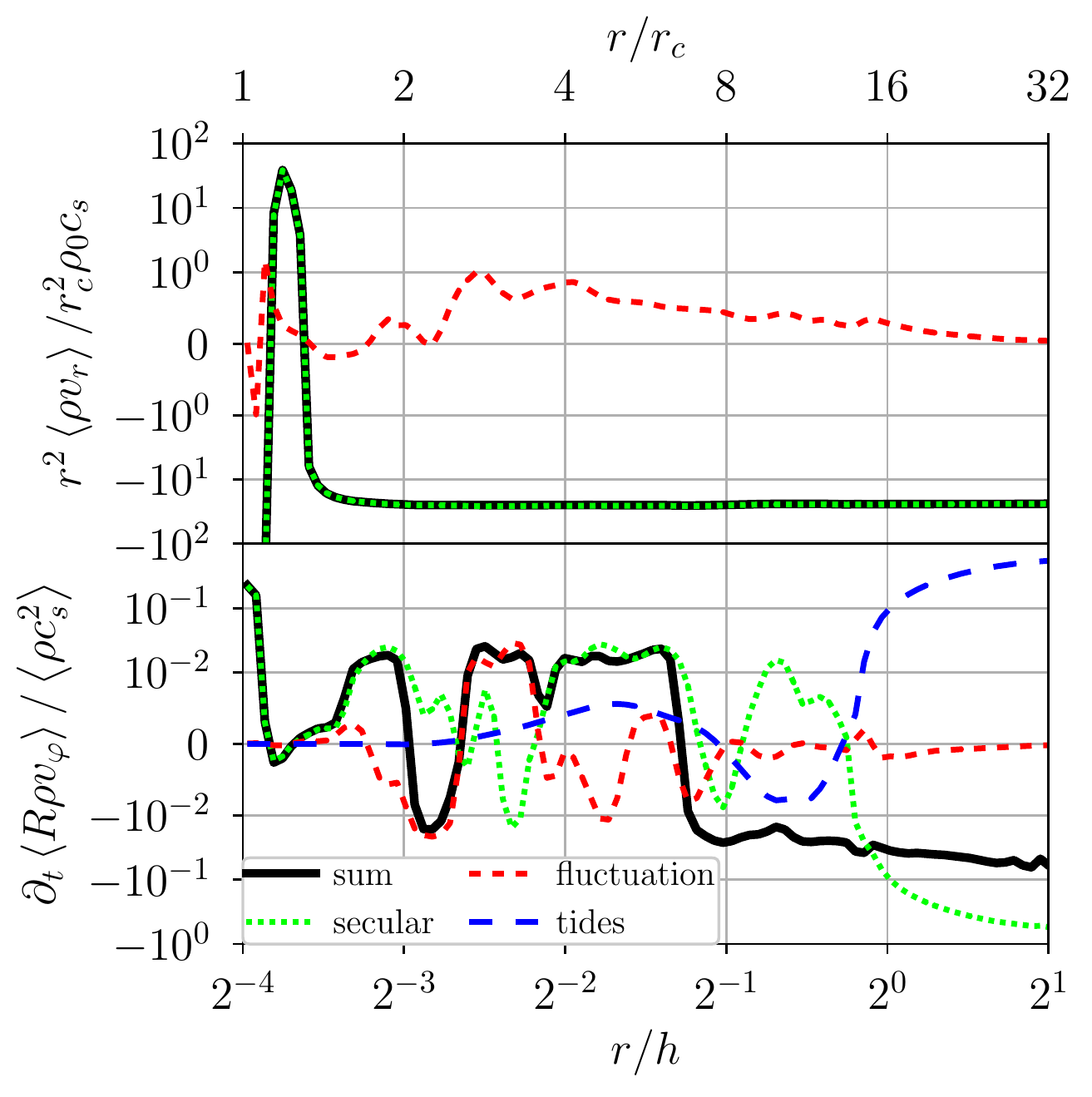}
\caption{\emph{Upper panel}: Time and shell-averaged radial mass flux in run \texttt{H16B16} (solid black), decomposed into secular (dotted green) and fluctuating (dashed red) fluxes. \emph{Lower panel}: Average torques in the same run, decomposed into the mean component (dotted green), fluctuating component (dashed red) and gravitational torques (dashed blue). \label{fig:i3dp16b16_SH_masstorques}}
\end{center}
\end{figure}

The upper panel of \autoref{fig:i3dp16b16_SH_masstorques} shows that there is a net mass flux accreting through the envelope. As $\partial_r\left[r^2\rho v_r\right]\approx 0$, this accretion flow is steady in time except in a narrow shell on top of the core where mass accumulates. Sampling the density and velocity distributions over time, we can separate the secular and fluctuating components of the radial mass flux. The mass $m(r,t)$ inside a sphere of radius $r$ evolves on long (secular) timescales as
\begin{equation}
\frac{\partial m}{\partial t} = - 4\pi r^2 \left(\overline{\rho}\cdot \overline{v_r} + \overline{\rho' v_r'} \right).
\end{equation}
The mass inside $r<h$ increases almost linearly in time, with a volume averaged density growing as $\brac{\rho}/\rho_0 \sim 1.77 \,\Omega t/ 2\pi$. At this rate, it would take approximately $300$ orbits to reach the mass of the corresponding hydrostatic envelope. As our simulations are substantially shorter, we do not reach the stage of the envelope being in nearly hydrostatic equilibrium, although we do detect a reduction of the inflow velocity in time. We recall that without self-gravity or radiative transfer, there is no characteristic density scale, so the envelope mass is proportional to the background disc density $\rho_0$ regardless of the core mass. The rate at which the envelope grows matches the surface-integrated mass flux to better than $10^{-2}$ accuracy, so there is no noticeable mass loss through the inner radial boundary (surface of the core).

The mass flux $\brac{\overline{\rho' v_r'}}$ estimated from the density and velocity fluctuations is positive at every radius. This fluctuating mass flux does not only come from turbulent motions, but also from the fact that the `secular' mass flux evolves over the time-averaging interval. Indeed, as the envelope grows more massive ($\brac{\rho'}>0$), the accretion velocity tends to decrease ($\brac{v_r'}>0$), so the correlation $\brac{\overline{\rho' v_r'}}$ has a non-zero secular component that is generally positive. By narrowing the time-averaging interval, we can reduce the amplitude of this secular contribution and still estimate the amplitude of turbulent motions. We found that restricting the averaging interval to three orbits is sufficient to extract a statistically meaningful turbulent mass flux. Doing so, we find that the turbulent mass flux is smaller than the secular component by more than a factor of $10$ at every radius, i.e. negligible with respect to mass accretion. This is similar to what was found in 2D case in the analogous run with $B=H=16$ (see Paper I).

We measure the contribution of rotational support in the envelope of run \texttt{H16B16} by computing the maximal value of $v_{\varphi}/\vK \approx 38$ per cent in the equatorial plane. To become more massive (more pressure-supported), the envelope must lose momentum. The lower panel of \autoref{fig:i3dp16b16_SH_masstorques} shows the net torque exerted on the fluid inside the envelope. The net torque oscillates around zero; its maximal amplitude corresponds to a momentum removal timescale of $\sim 10^2$ orbits. Mass accumulates only near the core surface, so accretion is likely allowed by momentum losses in its boundary layer. In the midplane, the core surface can exchange angular momentum with the surrounding fluid by both turbulent and numerical diffusion. Dedicated boundary layer simulations would be required to disentangle these two effects \citep[e.g.,][]{belyaev13a}. In the polar regions, the supersonic inflows terminate in shocks (see \autoref{fig:i3dp16b16_PLm_rhodv}), and these shocks accurately delimit the region where mass accumulates. Because the gas is taken to be isothermal, there is no heat generation at shocks, so energy is effectively dissipated there. Because the shocked gas comes from high latitudes, it carries virtually no angular momentum. The isothermal shocks thus allow mass to accumulate deep inside the envelope without any need for angular momentum extraction. 

%%%%%%%%%%%%%%%%%%%%%%%%%%%%%%%%%%%%%%%%%%%%%%%%%%

\subsection{Time variability}

%%%%%%%%%%%%%%%%%%%%%%%%%%%%%%%%%%%%%%%%%%%%%%%%%%

\begin{figure}%[H]
\begin{center}
\includegraphics[width=1.0\columnwidth]{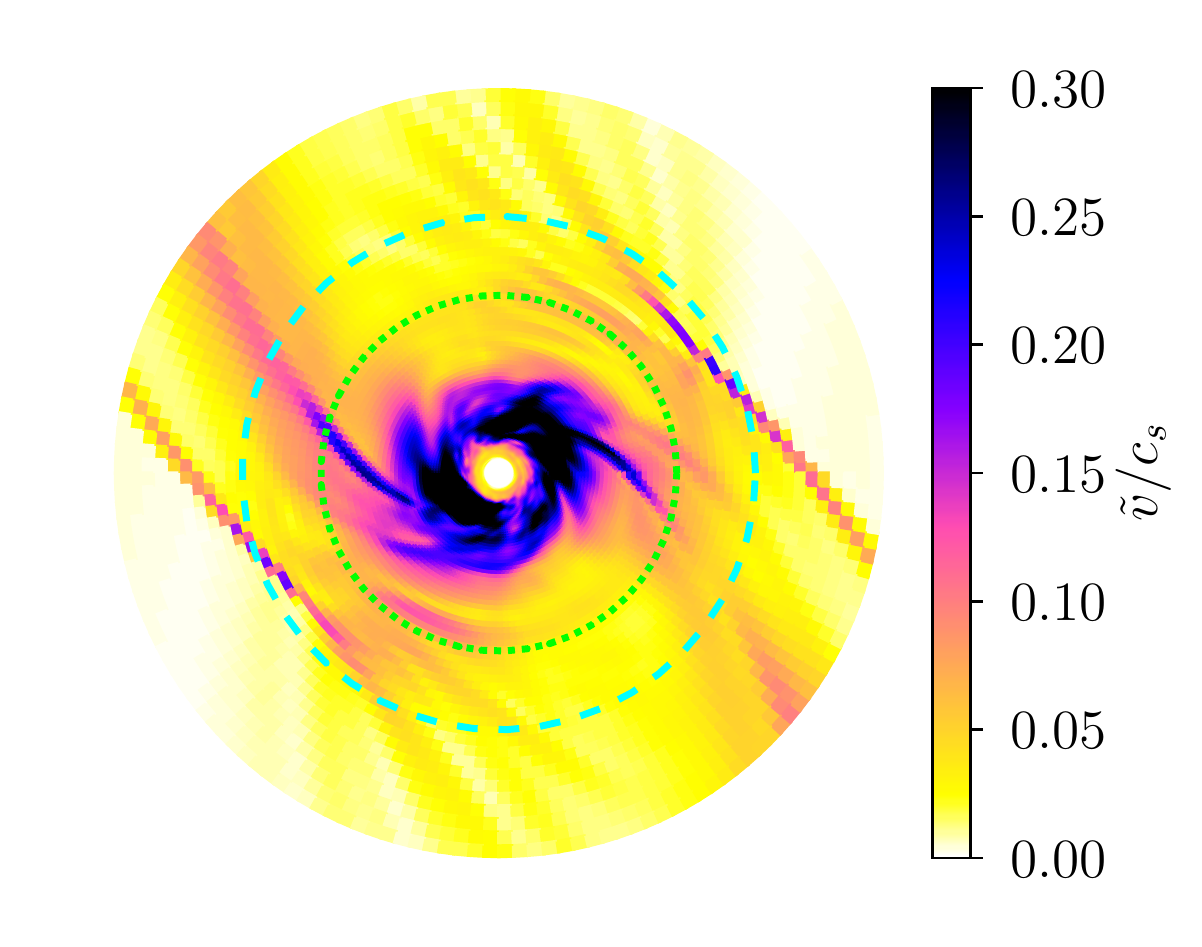}
\caption{Time-averaged turbulent Mach number in the equatorial plane of run \texttt{H16B16}. The dashed cyan and dotted green circles respectively mark the pressure scale $h$ and the Hill radius $\rH$. 
\label{fig:i3dp16b16_EQt_Vmm}}
\end{center}
\end{figure}

\begin{figure}%[H]
\begin{center}
\includegraphics[width=1.0\columnwidth]{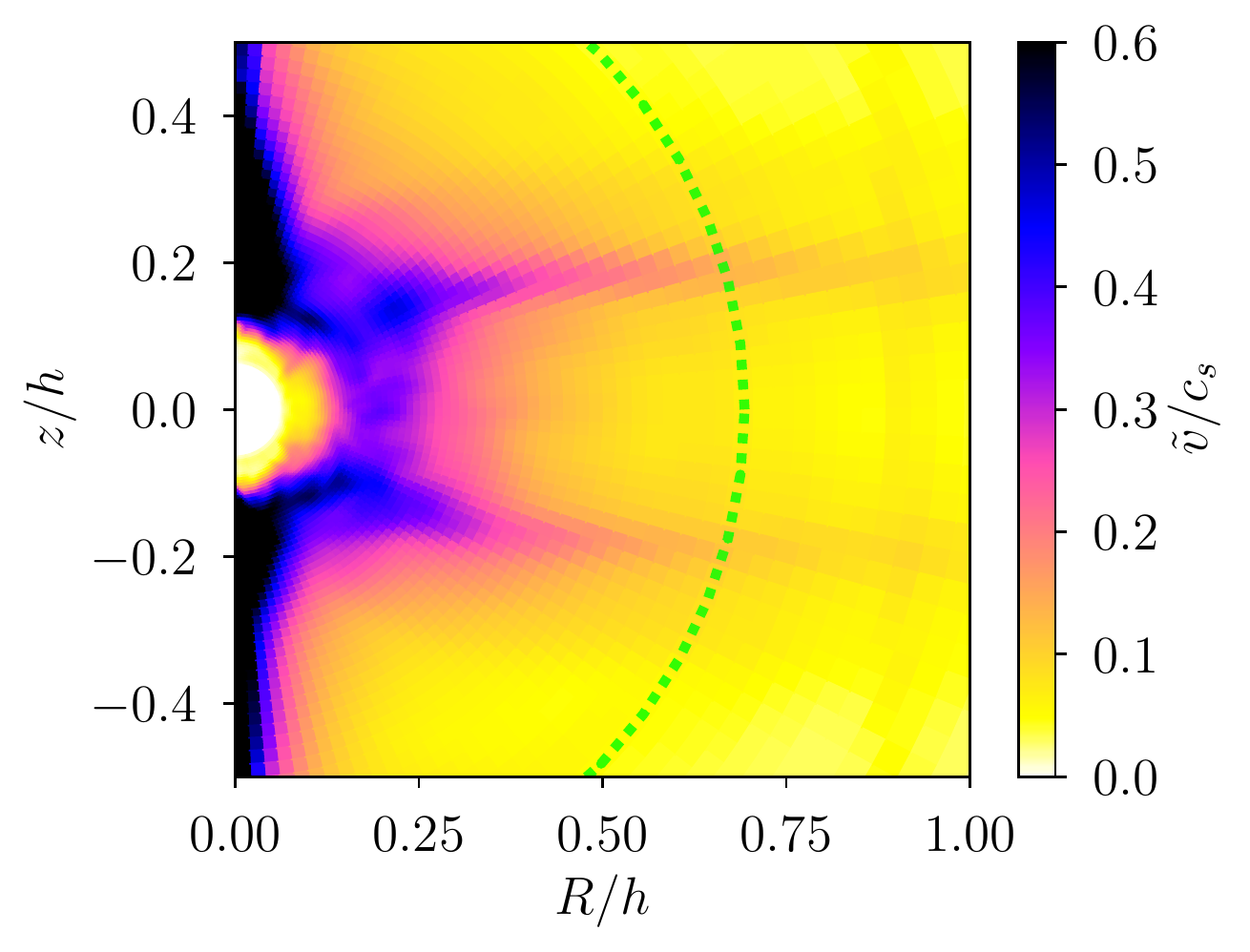}
\caption{Time and azimuthally-averaged turbulent Mach number in the poloidal plane of run \texttt{H16B16}. The Hill radius $\rH$ is marked by a dotted green circle. The color scale is truncated for a better contrast but $\tilde{v}/c_s \lesssim 20$ near the polar axis. See text for details.
\label{fig:i3dp16b16_PLm_Vmm}}
\end{center}
\end{figure}

Velocity fluctuations can enhance the mixing of the gas from different regions of the envelope, as well as the exchange of gas between the disc and the envelope. Even without inducing a net mass flux, turbulent diffusion could affect the thermal structure of an envelope by homogenizing its entropy toward the background disc value. We start adressing this issue by examining the amplitude of velocity fluctuations in our isothermal simulations. 

The envelope of run \texttt{H16B16} features small-scale variability on short timescales. We estimate the turbulent Mach number $\mathrm{M} \equiv \tilde{v}/c_s$ from the velocity fluctuations $\tilde{v}^2 \equiv \overline{v'\cdot v'}$, and represent its equatorial and poloidal distributions on \autoref{fig:i3dp16b16_EQt_Vmm} and \autoref{fig:i3dp16b16_PLm_Vmm} respectively.

In the equatorial plane, \autoref{fig:i3dp16b16_EQt_Vmm} reveals variability in the entire envelope. The turbulent Mach number is maximal inside $r/\rH<1/2$ with $\mathrm{M} \approx 30$ per cent. There is a thin shell directly on top of the core where the turbulent Mach number drops below $10$ per cent. None of our isothermal simulations show supersonic turbulence in the equatorial plane: the turbulent Mach number always saturates below $50$ per cent. 

\autoref{fig:i3dp16b16_PLm_Vmm} shows that the velocity fluctuations are stronger away from the midplane. The variability at $z/h\approx \pm 0.2$ is excited by helical motions of the fluid as shown on Figure 10 of \cite{fung15}; these sources of variability are not axisymmetric. Variability is mostly excited in the polar regions, with velocity fluctuations reaching supersonic amplitudes. The turbulent Mach number goes as high as $\mathrm{M} \approx 1.2$ at the polar shocks, half the amplitude of the time-averaged inflow Mach number. The sonic surface marked on \autoref{fig:i3dp16b16_PLm_rhodv} actually fluctuates on short time and spatial scales near the core. In the post-shock region surrounding the core, the turbulent Mach drops from $\mathrm{M}\approx 1$ at $z=2~r_c$ down to $\mathrm{M}\approx 10^{-2}$ at the core surface. Numerical dissipation might damp the turbulent motions at the core surface, but most of the momentum dissipation has already occured through the polar shocks. The only simulation run maintaining subsonic variability in the entire envelope is \texttt{H16B8}, with a turbulent Mach number $\mathrm{M} \lesssim 6$ per cent on the polar axis. 

%%%%%%%%%%%%%%%%%%%%%%%%%%%%%%%%%%%%%%%%%%%%%%%%%%

\subsection{Envelope recycling} 
\label{sec:envrecycle}

%%%%%%%%%%%%%%%%%%%%%%%%%%%%%%%%%%%%%%%%%%%%%%%%%%

We now look at the issue of the envelope recycling by exchange of gas between the disc and the envelope interior \citep{ormel2}. In Paper I we used passive tracers to diagnose fluid exchanges between the different parts of the flow; this is a method we resort to in this work as well. In our fiducial run the flow can be considered as being in a quasi-steady state after ten orbits of integration time. At this point we inject a tracer fluid with constant concentration $n=1$ inside the Bondi sphere. The tracer is passively advected by the flow according to $\partial_t n + v\cdot\nabla n = 0$, or equivalently
\begin{equation} \label{eqn:dttr1}
\partial_t \left( n \rho \right) + \nabla\cdot\left( n \rho v\right) = 0
\end{equation}
for the traced mass. We follow the evolution of the traced fluid over ten additional orbits since tracer injection. 

\begin{figure}%[H]
\begin{center}
\includegraphics[width=1.0\columnwidth]{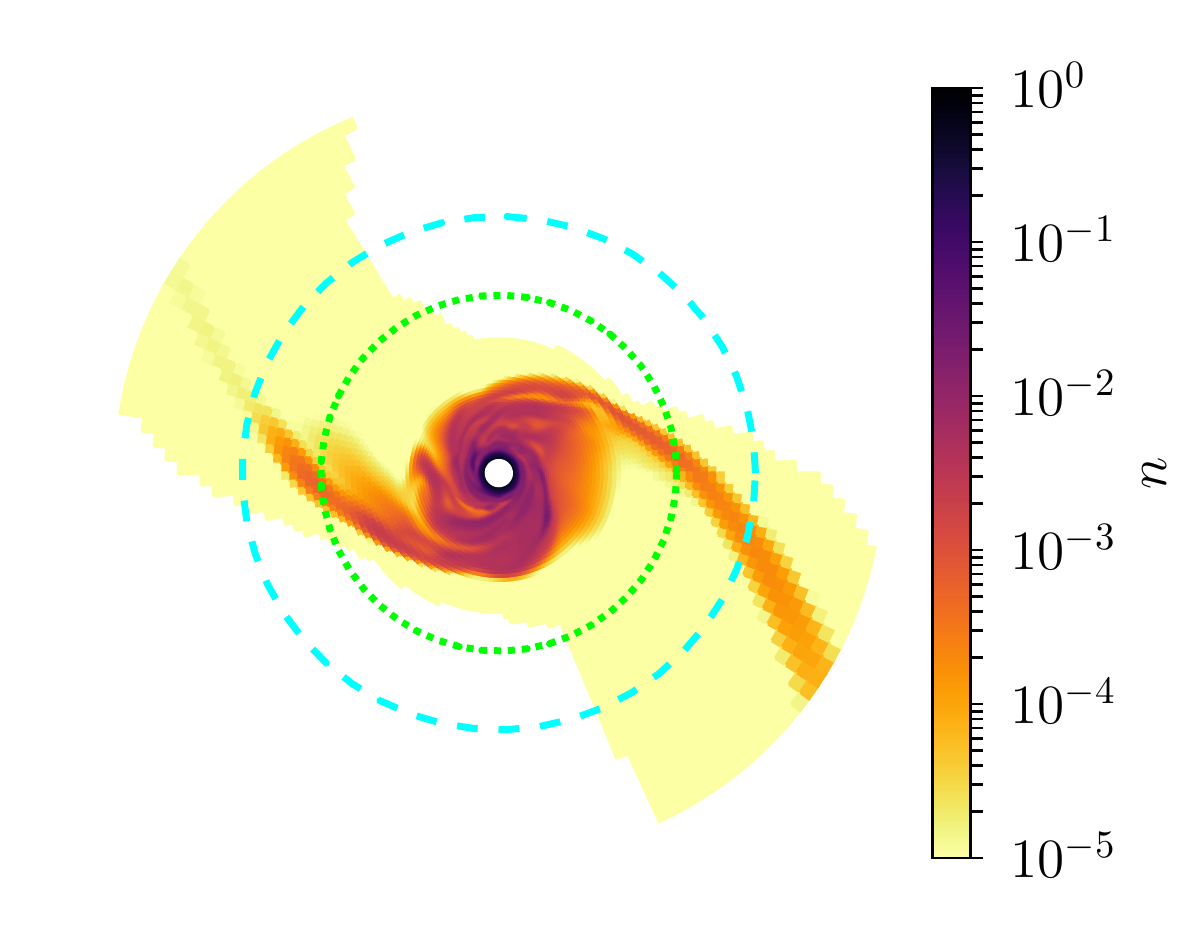}
\caption{Equatorial distribution of tracer concentration ten orbits after its injection in the Bondi sphere in run \texttt{H16B16}. The dashed cyan and dotted green circles respectively mark the pressure and Hill radius. 
\label{fig:i3dp16b16_EQt_tr1}}
\end{center}
\end{figure}

\autoref{fig:i3dp16b16_EQt_tr1} shows the tracer distribution in the equatorial plane of run \texttt{H16B16} ten orbits after tracer injection. The remaining tracer is concentrated inside $r\lesssim \rH/2$, where the time-averaged flow features circular streamlines orbiting the core (see \autoref{fig:i3dp16b16_EQt_tot}). The tracer concentration displays a spiral pattern offset with respect to the spiral density waves. These spiral streams do not result from gas compression: they reveal how the traced fluid is channelled out of the envelope by gas flowing from high latitudes. The tracer concentration is still of order unity at the surface of the core. 

\begin{figure}%[H]
\begin{center}
\includegraphics[width=1.0\columnwidth]{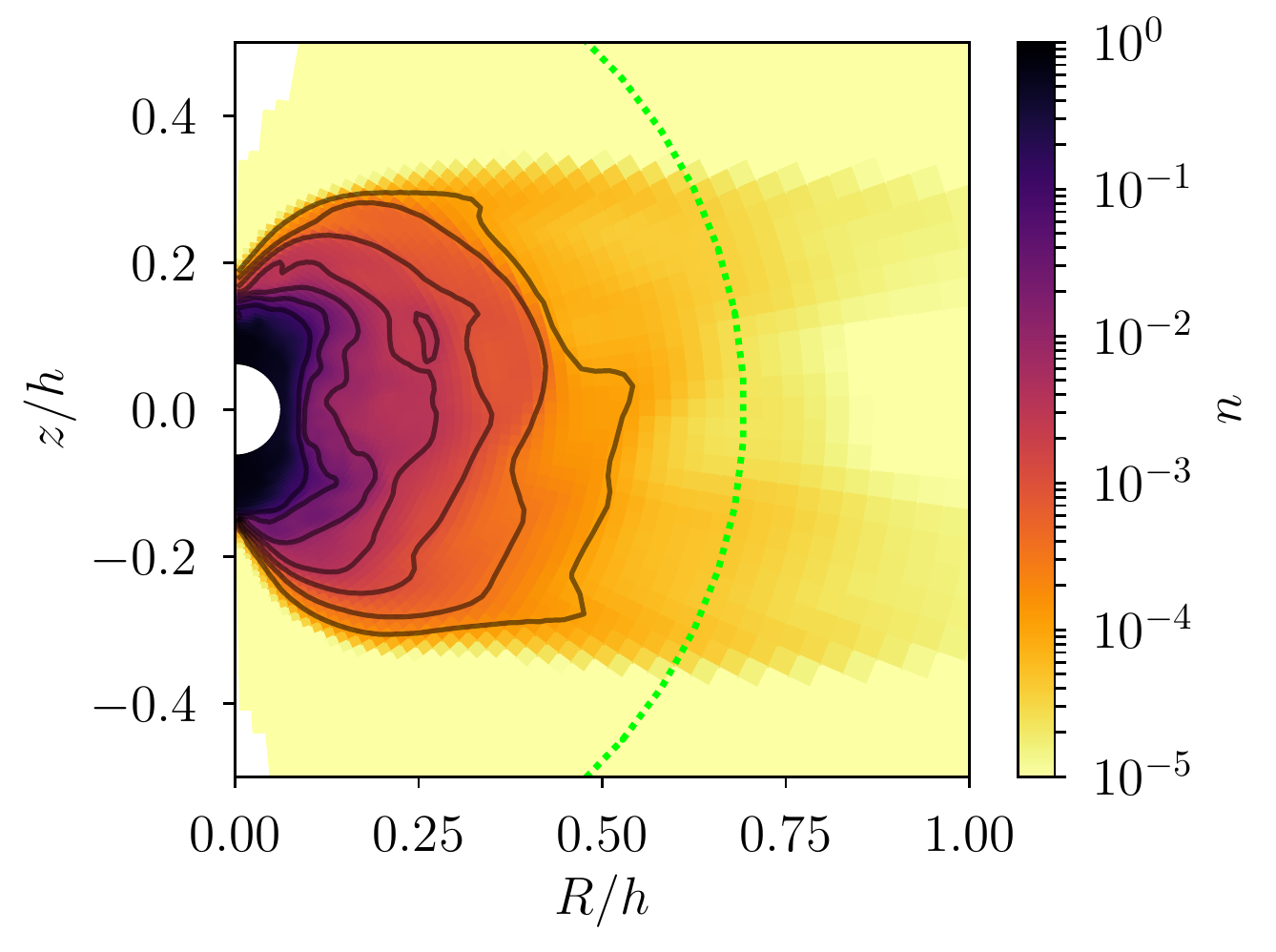}
\caption{Azimuthally averaged tracer concentration $\brac{n}_{\varphi}$ in run \texttt{H16B16} ten orbits after tracer injection (color scale and iso-contours). The Hill radius $\rH$ is marked by a dotted green circle. \label{fig:i3dp16b16_PLm_tr1}}
\end{center}
\end{figure}

\autoref{fig:i3dp16b16_PLm_tr1} shows the average tracer distribution in the poloidal plane of run \texttt{H16B16}. The iso-concentration contours form lobes restricted to $\vrac{z/h} \lesssim 0.3$ and pinched toward the core on the polar axis. The traced fluid is never transported high above the midplane, it is only depleted there. The tracer concentration has barely decreased from its initial value near the polar caps of the core. In this region where $R\lesssim r_c$, recycling is inefficient up to the polar shocks at $\vrac{z} \approx 2.4~r_c$, inside of which radial motions are largely suppressed. 

\begin{figure}%[H]
\begin{center}
\includegraphics[width=1.0\columnwidth]{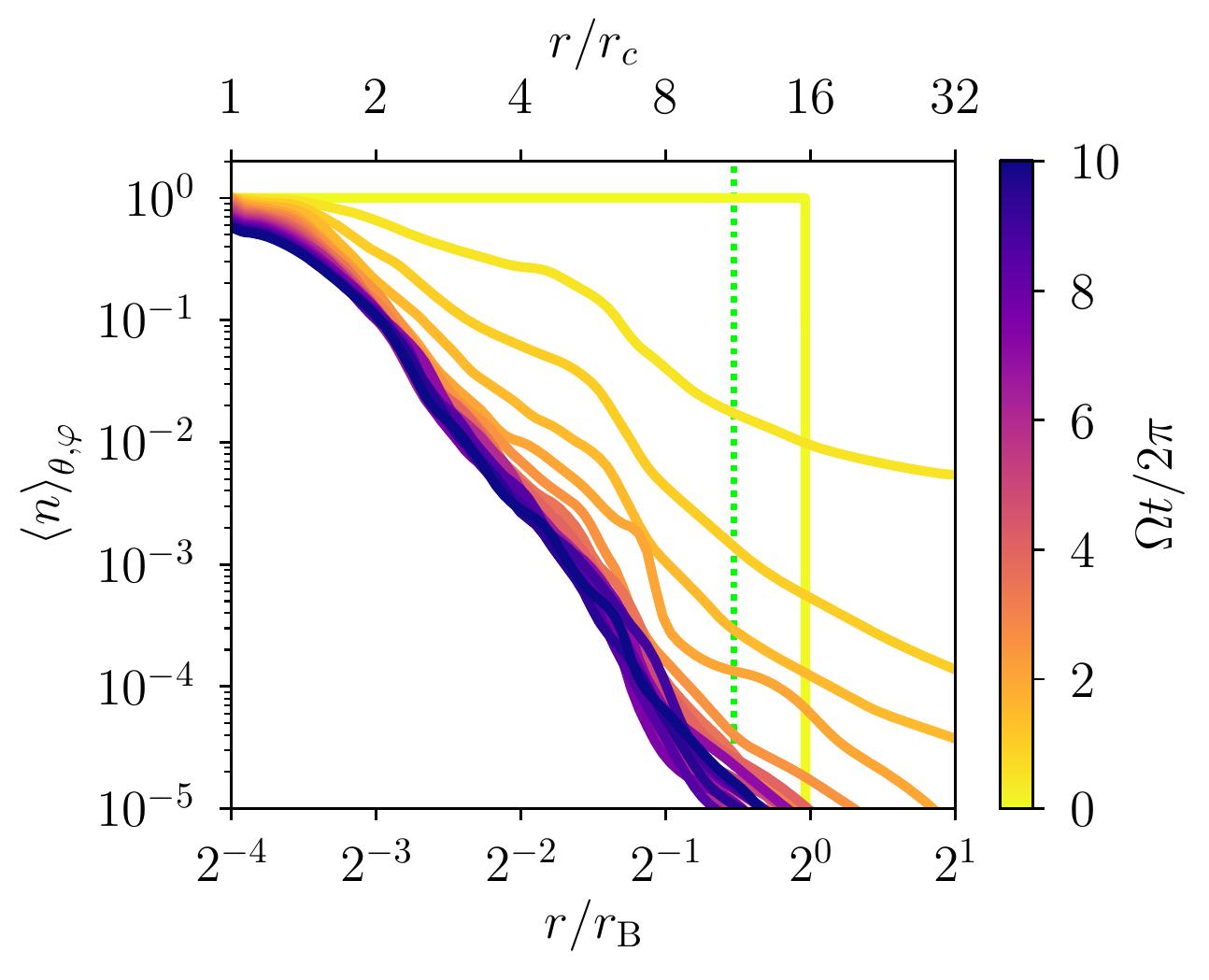}
\caption{Evolution of the shell-averaged tracer concentration profile $\langle n(r)\rangle$ over time (curves and color scale) since tracer injection in the Bondi sphere of run \texttt{H16B16}. The dotted vertical green line marks the Hill radius $\rH$. \label{fig:i3dp16b16_SH_tsv}}
\end{center}
\end{figure}

\autoref{fig:i3dp16b16_SH_tsv} shows the evolution over time of the shell-averaged tracer concentration in run \texttt{H16B16}. It takes approximately two orbits to reach a quasi-steady concentration profile. Compared to two-dimensional simulations, there is no plateau in the concentration profile: the tracer concentration is substantially reduced by envelope mixing down to the core surface at $r\approx r_c$. The average tracer concentration keeps slowly decreasing inside $r\lesssim 2~r_c$. The radially cumulated tracer mass decreases with an e-folding decay timescale going from $10$ to $30$ orbits over the duration of the simulation. This tracer dispersal timescale is not adequately defined because the geometry of the tracer distribution keeps evolving in time. Different regions of the envelope are recycled on different timescales: material in the midplane is evacuated faster than material near the poles, and the flow does not efficiently homogenize the tracer distribution near the core. We discuss the use of another dispersion timescale in \autoref{sec:recycling}.

%%%%%%%%%%%%%%%%%%%%%%%%%%%%%%%%%%%%%%%%%%%%%%%%%%
%%%%%%%%%%%%%%%%%%%%%%%%%%%%%%%%%%%%%%%%%%%%%%%%%%

\section{Results: parameter exploration}
\label{sect:res_vary}

%%%%%%%%%%%%%%%%%%%%%%%%%%%%%%%%%%%%%%%%%%%%%%%%%%

We now explore the sensitivity of the different diagnostics presented in the previous section when varying the core mass and size via the parameters $H$ and $B$. 

According to equation (\ref{eq:mth-rel}), $B/H=m_c / m_{\rm th}$, so that increasing (decreasing) $B$ and $H$ in the same proportions corresponds to decreasing (increasing) the core radius without changing its mass relative to the thermal mass or the mass of its host star. On the other hand, changing $B$ alone while keeping $H$ fixed corresponds to varying the mass of the core without changing its size relative to the disc scale height $h$. 

\begin{figure*}
\begin{tabular}{cc}
\subfloat[\texttt{H16B8}, $m_c/m_{\mathrm{th}}=1/2$]{\includegraphics[width=1.0\columnwidth]{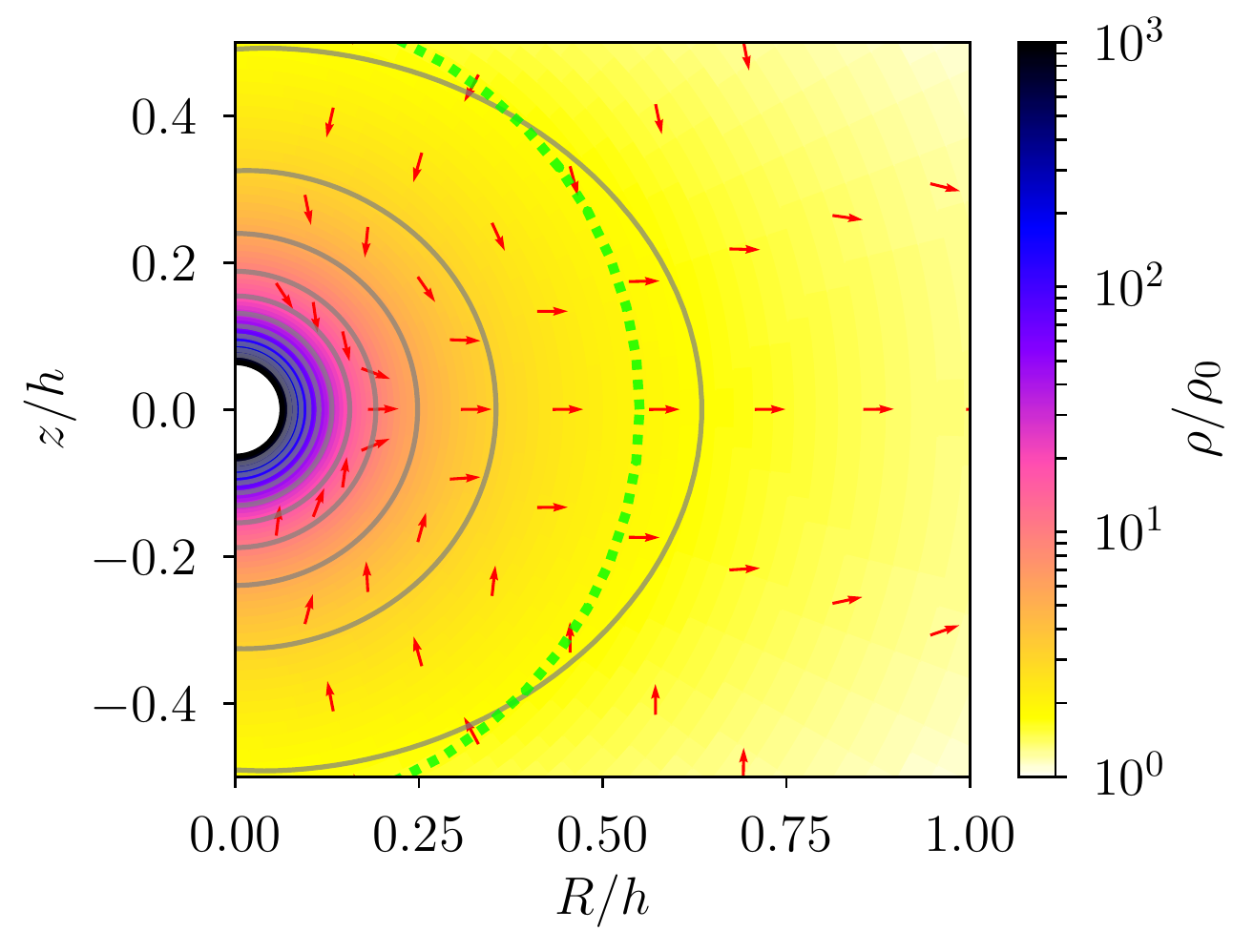}} &
\subfloat[\texttt{H16B16}, $m_c/m_{\mathrm{th}}=1$]{\includegraphics[width=1.0\columnwidth]{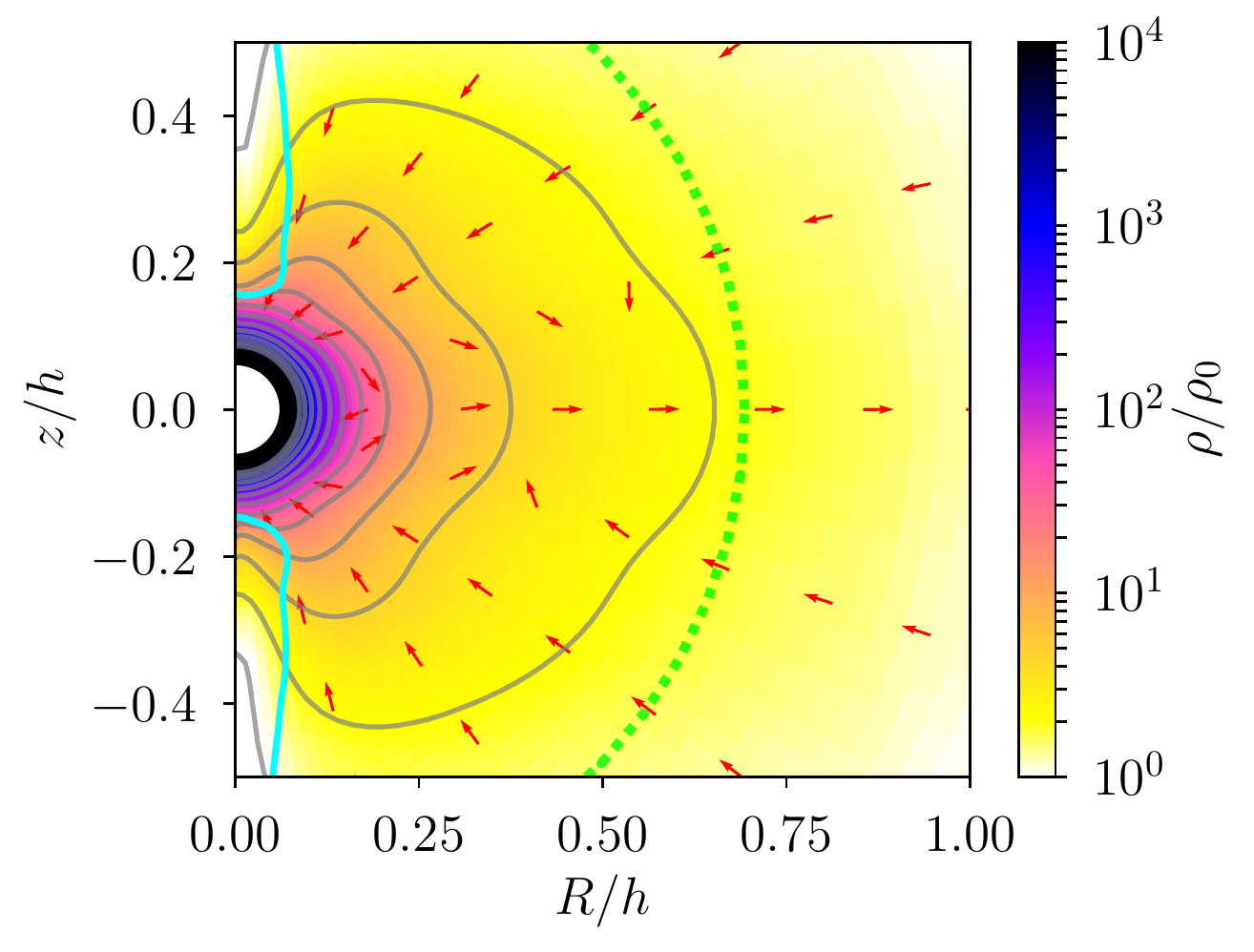}} \\
\subfloat[\texttt{H16B32}, $m_c/m_{\mathrm{th}}=2$]{\includegraphics[width=1.0\columnwidth]{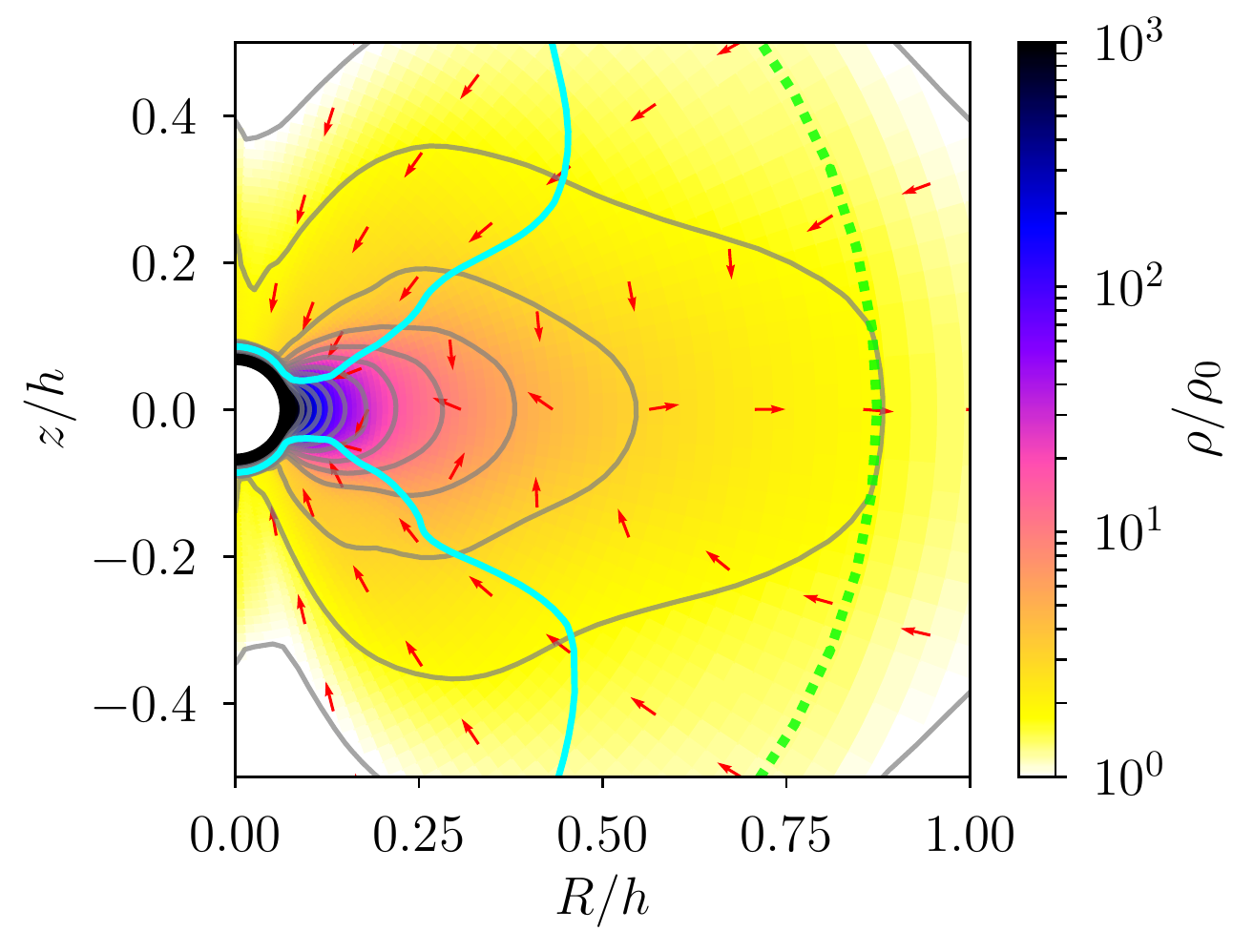}} &
\subfloat[\texttt{H16B64}, $m_c/m_{\mathrm{th}}=4$]{\includegraphics[width=1.0\columnwidth]{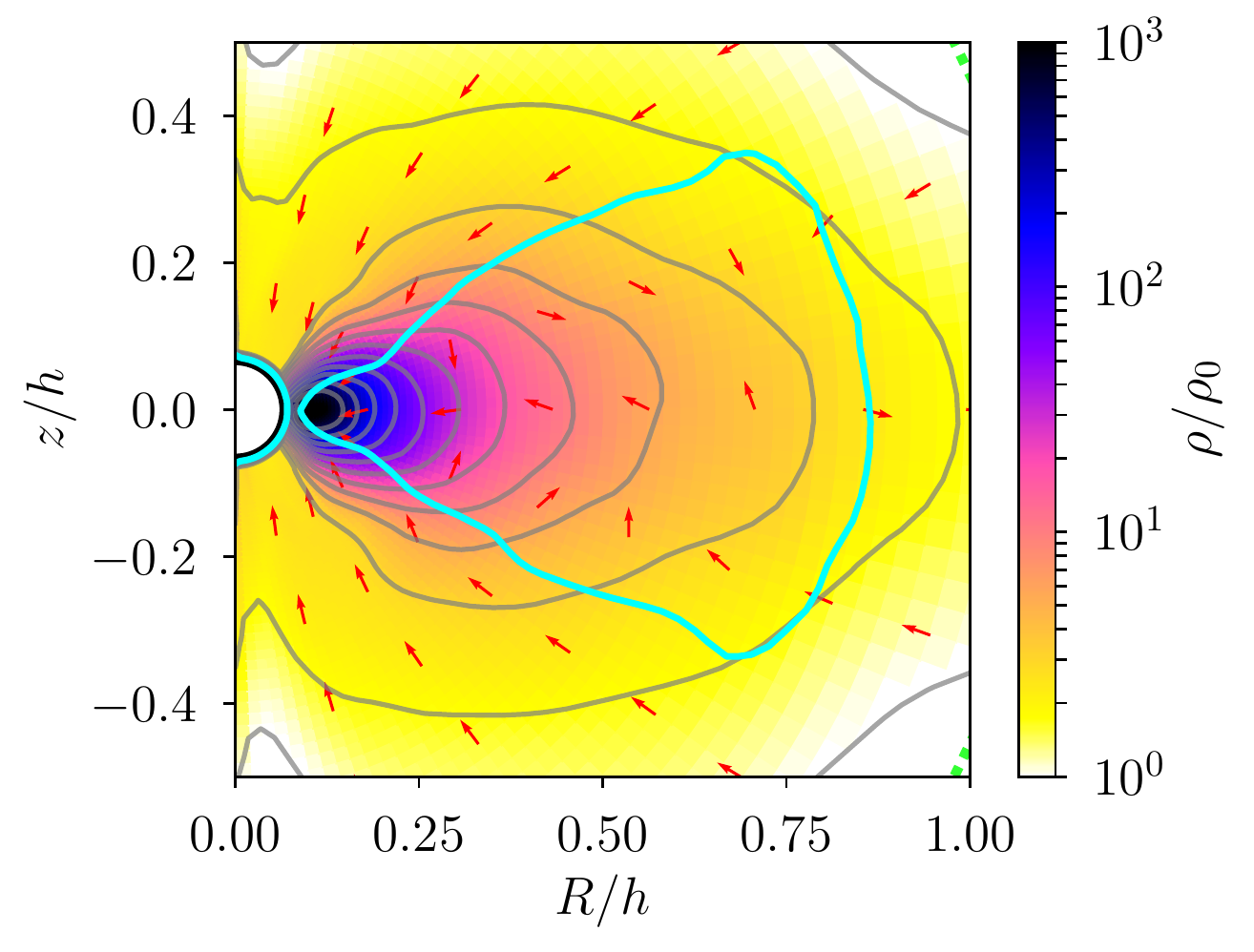}} \\
\subfloat[\texttt{H32B16}, $m_c/m_{\mathrm{th}}=1/2$]{\includegraphics[width=1.0\columnwidth]{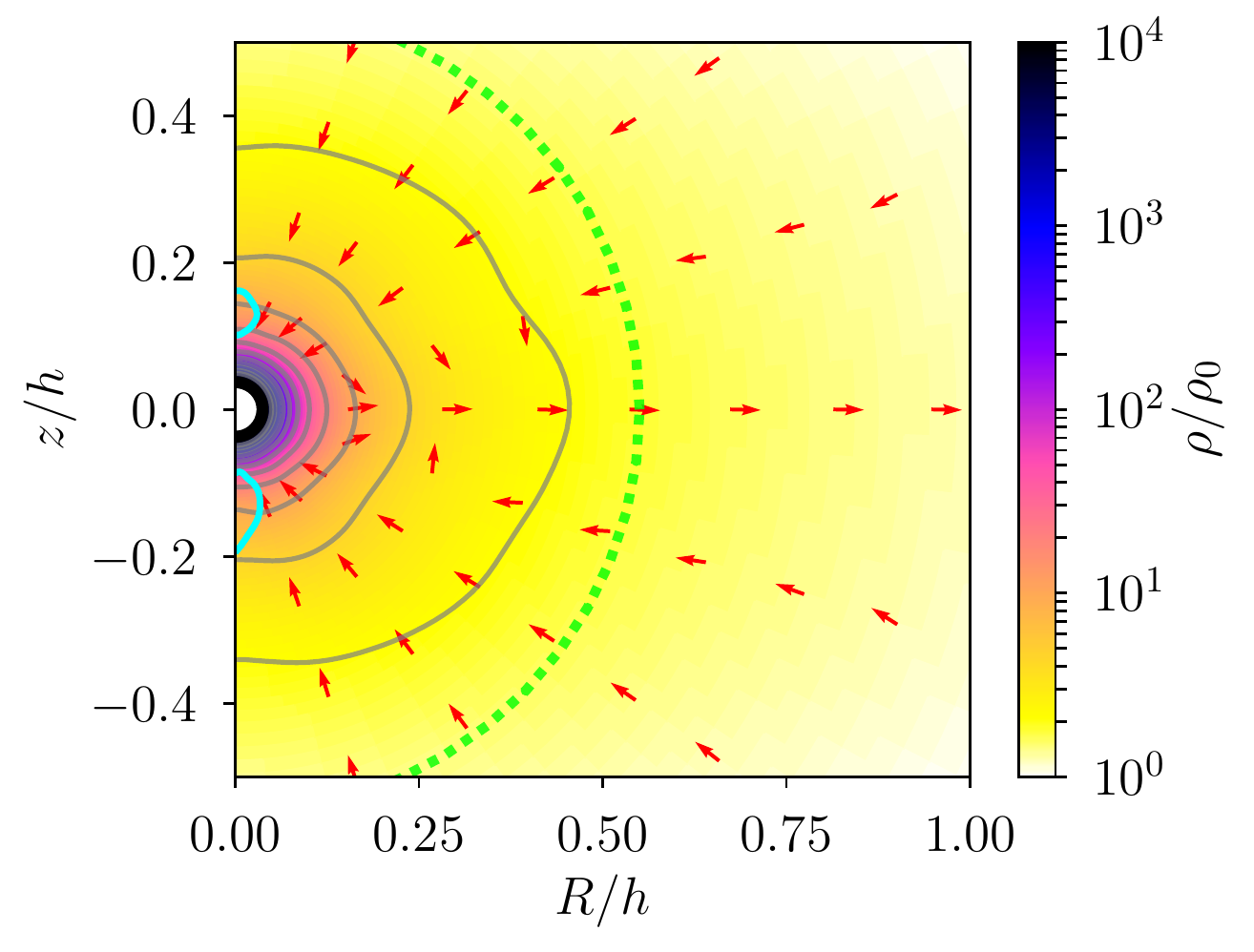}} &
\subfloat[\texttt{H32B32}, $m_c/m_{\mathrm{th}}=1$]{\includegraphics[width=1.0\columnwidth]{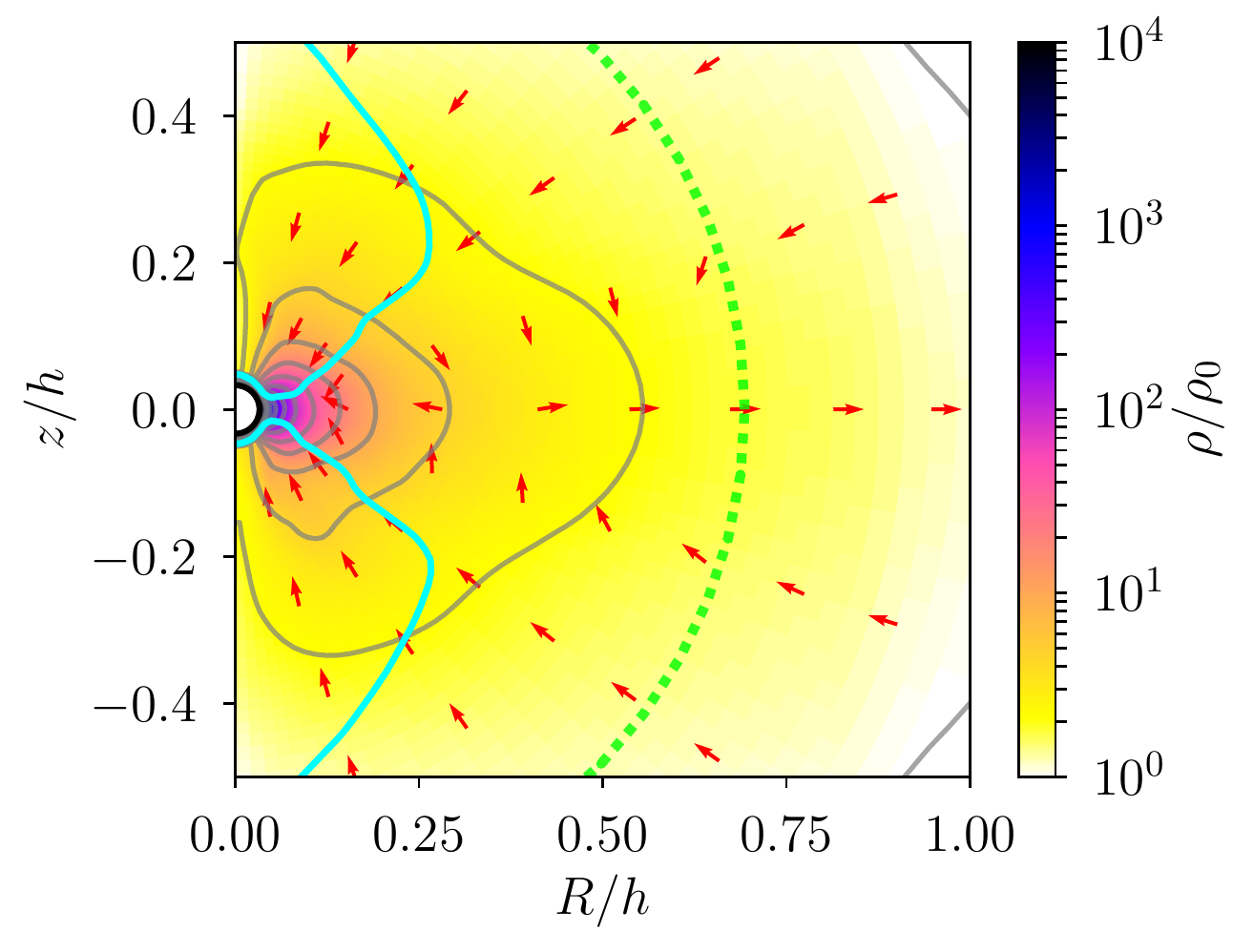}}
\end{tabular}
\caption{Time and azimuthally-averaged density (color map and grey contours) and poloidal mass flux (red arrows, orientation only) in the meridional plane of runs \texttt{H16B8} (a), \texttt{H16B16} (b, reference run); \texttt{H16B32} (c), \texttt{H16B64} (d), \texttt{H32B16} (e) and \texttt{H32B32} (f). The thick cyan line near the poles marks the sonic surface of the poloidal velocity, and the Hill radius $\rH$ is marked by a dotted green circle.} \label{fig:PLm_rhodv}
\end{figure*}

\autoref{fig:PLm_rhodv} gathers the poloidal maps of time and azimuthally-averaged density and mass flux in our series of simulations listed in Table \ref{tab:recap}. The top four panel have $H=16$ and increasing values of $B$, while the bottom two panels have $H=32$. We proceed to compare them according to the mass and size of the core. 

%%%%%%%%%%%%%%%%%%%%%%%%%%%%%%%%%%%%%%%%%%%%%%%%%%

\subsection{Varying the core mass}
\label{sect:core_mass_var}

%%%%%%%%%%%%%%%%%%%%%%%%%%%%%%%%%%%%%%%%%%%%%%%%%%

We start by comparing simulations having the same $H$ to examine the influence of the core mass on the envelope properties. A series of runs with $H=16$ and $B=8$, $16$, $32$ and $64$ shown in \autoref{fig:PLm_rhodv}a-d spans $m_c/m_{\rm th}=0.5$, $1$, $2$, and $4$, thus exploring the transition from low mass cores (previously studied by, e.g., \citealt{ormel2}) to high mass cores \citep[e.g.,][]{bate03,szulagyi16} using a single numerical setup.

First, we look at the density distribution around the core. At the lowest mass sampled (\texttt{H16B8}, $m_c=0.5m_{\rm th}$), the iso-density contours are rather close to being spherical with only a small amount of oblateness that increases with radius, see \autoref{fig:PLm_rhodv}a. This deviation from spherical symmetry is caused by a weak rotational support in the envelope, which will be examined in more detail in \autoref{sect:rot_support}. As $m_c$ is increased, the density distribution becomes compressed toward the midplane, see the fiducial run \texttt{H16B16} with $m_c=m_{\rm th}$ on \autoref{fig:PLm_rhodv}b. As $m_c$ increases further to $m_c=2~m_{\rm th}$ (\texttt{H16B32}, \autoref{fig:PLm_rhodv}c), more mass becomes concentrated in an oblate overdensity near the midplane. Eventually, in the run \texttt{H16B64}, where the core is four times more massive than $m_{\rm th}$, the density contours are no longer focused on the core: there is a local density maximum at $r\approx 1.8r_c$ in the midplane, disjoint from the core surface. This is only possible if rotational support dominates the radial momentum balance, i.e. $v_{\varphi}/\vK \gtrsim 1$. These are unambiguous signatures of a forming circumplanetary disc. 

Next, we look at the poloidal flow pattern in this set of simulations. As mentioned in \autoref{sect:flow_fid}, the gas inflow towards the core occurs at high latitudes regardless of the core mass. Some of this flow gets accreted by the core, while a fraction gets diverted towards the midplane and is then funneled outwards, returning to the bulk of the disc for all cores with $m_c\lesssim 2m_{\rm th }$. This circulation pattern featuring an outflow of gas in the midplane is very different from the picture of equatorial accretion established in the 2D case \citep[e.g.,][]{kley99}. The midplane gas outflow present in the circumplanetary region for all planetary masses is likely to oppose or even prevent the accretion of small particles (pebbles) by the core \citep{Kuwahara}.

The situation is a little different only for the most massive core of run \texttt{H16B64} ($m_c=4m_{\rm th}$), as can be see in Figure \ref{fig:PLm_rhodv}d. Here the inflow has too much angular momentum to reach the core and is deflected toward the midplane by the `centrifugal barrier' even at intermediate latitudes. The resultant pileup of mass in a disc-like structure prevents the high-latitude flow from being reoriented outward in the equatorial plane. Instead, the gas keeps piling up in this circumplanetary disc whose mass increases on orbital timescales. We find that in this simulation the volume-averaged density (inside $r<h$) grows linearly in time as $\brac{\rho}/\rho_0 \sim 7.0\, \Omega t/2\pi$. Although this is four times faster than in the fiducial run \texttt{H16B16} (see \autoref{sec:fidqsstate}), the time required for the envelope to reach the mass corresponding to hydrostatic equilibrium is now much larger\footnote{The mass of a hydrostatic envelope grows as $\int_{r_c}^{h} 4\pi\rho_0 r^2 e^{\rB/r} \dd r$, which is faster than exponential with the core mass.}. 

One can also see that variation of $m_c$ strongly affects the characteristics of the near polar gas inflow. In the low-mass case of run \texttt{H16B8} (\autoref{fig:PLm_rhodv}a) the flow remains subsonic in the whole domain. However, as pointed out in \autoref{sect:flow_fid}, a transsonic surface appears around polar axis in the fiducial run \texttt{H16B16}. The inflow along this axis gets eventually arrested in a standing shock not too far from the core surface, and mass accumulates there. As $m_c$ increases, the sonic surface extends to lower latitudes, and the standing shock at the poles moves closer to the core surface. Finally, in the run \texttt{H16B64}, supersonic accretion streams span a wide cone about the polar axis. At high latitudes, the inflow reaches Mach $\mathrm{M}\approx 9.4$ and shocks at $z\approx 1.3~r_c$, nearly on the core surface. The post-shock medium is only barely resolved with seven grid cells in this case. The flow is subsonic only very near the core surface and also in the (detached) thick torus around the forming circumplanetary disc.

The strength of the polar shocks, i.e. the jump in Mach number $\mathrm{M}$, can roughly be estimated by assuming that the flow encounters the shock after falling freely from infinity. At the surface of the core, the corresponding Mach number would be $\mathrm{M}=(2Gm_c/r_s)^{1/2}/c_s=(2B)^{1/2}$. This estimate works reasonably well when correcting for the distance between the core surface and the shock front (see \autoref{sect:core_rad_var}). The maximum values of the polar Mach number are listed in Table \ref{tab:recap}.

%%%%%%%%%%%%%%%%%%%%%%%%%%%%%%%%%%%%%%%%%%%%%%%%%%

\subsection{Varying the core size}
\label{sect:core_rad_var}

%%%%%%%%%%%%%%%%%%%%%%%%%%%%%%%%%%%%%%%%%%%%%%%%%%

We now compare runs having the same ratio of $B/H$ to isolate the effect of the core size $r_c$ on the envelope properties.

\autoref{fig:PLm_rhodv}e represents the sub-thermal mass regime for a small core with $H=32$; it should be compared to \autoref{fig:PLm_rhodv}a, also featuring a core with $m_c=0.5~m_{\rm th}$ but with $H=16$. Despite the general similarities of the flow pattern (e.g., the orientation of the mass flux circulation), one notices several differences. First, a transsonic surface emerges in the polar regions of run \texttt{H32B16} even though it is in the sub-thermal mass regime. Second, the iso-density contours are mildly pinched near the midplane. This is explained by the significant level of rotational support in this case, with a maximum $v_\varphi/v_{\rm K}\approx 31$ per cent; in run \texttt{H16B8} this ratio does not exceed two percent. Third, the near-midplane outflow is now confined to a narrower range of latitudes around the equatorial plane of the simulation.

\autoref{fig:PLm_rhodv}f shows the poloidal density and mass flux maps of run \texttt{H32B32}, for which the core radius is twice smaller than in the reference run \texttt{H16B16} (see \autoref{fig:PLm_rhodv}b), while $m_c=m_{\rm th}$ for both. We notice several differences between the two cases. The density gradient, as measured by $\left\vert\partial \log \rho / \partial \log r\right\vert$ in the equatorial plane, is lower than in \texttt{H16B16}; to maintain radial momentum balance, the (maximum) fraction of rotational support $v_{\varphi}/\vK \approx 85$ per cent is consequently larger in run \texttt{H32B32} compared to the $38$ per cent of \texttt{H16B16}. Also, the sonic surface encompasses a wider cone around the core in run \texttt{H32B32}, i.e. a larger fraction of the inflow becomes supersonic before reaching the core. 

On the polar axis itself, the time-averaged flow of run \texttt{H32B32} reaches a Mach number $\mathrm{M}\approx 5.9$ before shocking on the densest parts of the envelope. The stationary polar shocks are located at $z/r_c\approx 1.7$ (14 grid cells), closer to the core than in the reference $H=16$ case. The free-fall velocity at this height is $\sqrt{2 G m_c / (1.7r_c)}$, corresponding to $\mathrm{M} \approx 6.1$ for a particle initially at rest away from the core. The agreement with the Mach number measured in simulation, $\mathrm{M}\approx 5.9$ (see Table \ref{tab:recap}), suggests that the inflow starts feeling a pressure gradient only near the shock, in the deepest layers of the envelope.

Despite these differences, the overall progression of the envelope properties with $m_c$ described in \autoref{sect:core_mass_var} for the case $H=16$ remains qualitatively the same for $H=32$, i.e. for twice smaller cores. Higher masses still result in stronger polar shocks, a higher degree of the rotational support of the envelope and more disc-like density distributions. 

%%%%%%%%%%%%%%%%%%%%%%%%%%%%%%%%%%%%%%%%%%%%%%%%%%

\subsection{Rotational support}
\label{sect:rot_support}

%%%%%%%%%%%%%%%%%%%%%%%%%%%%%%%%%%%%%%%%%%%%%%%%%%

\begin{figure}%[H]
\begin{center}
\includegraphics[width=1.0\columnwidth]{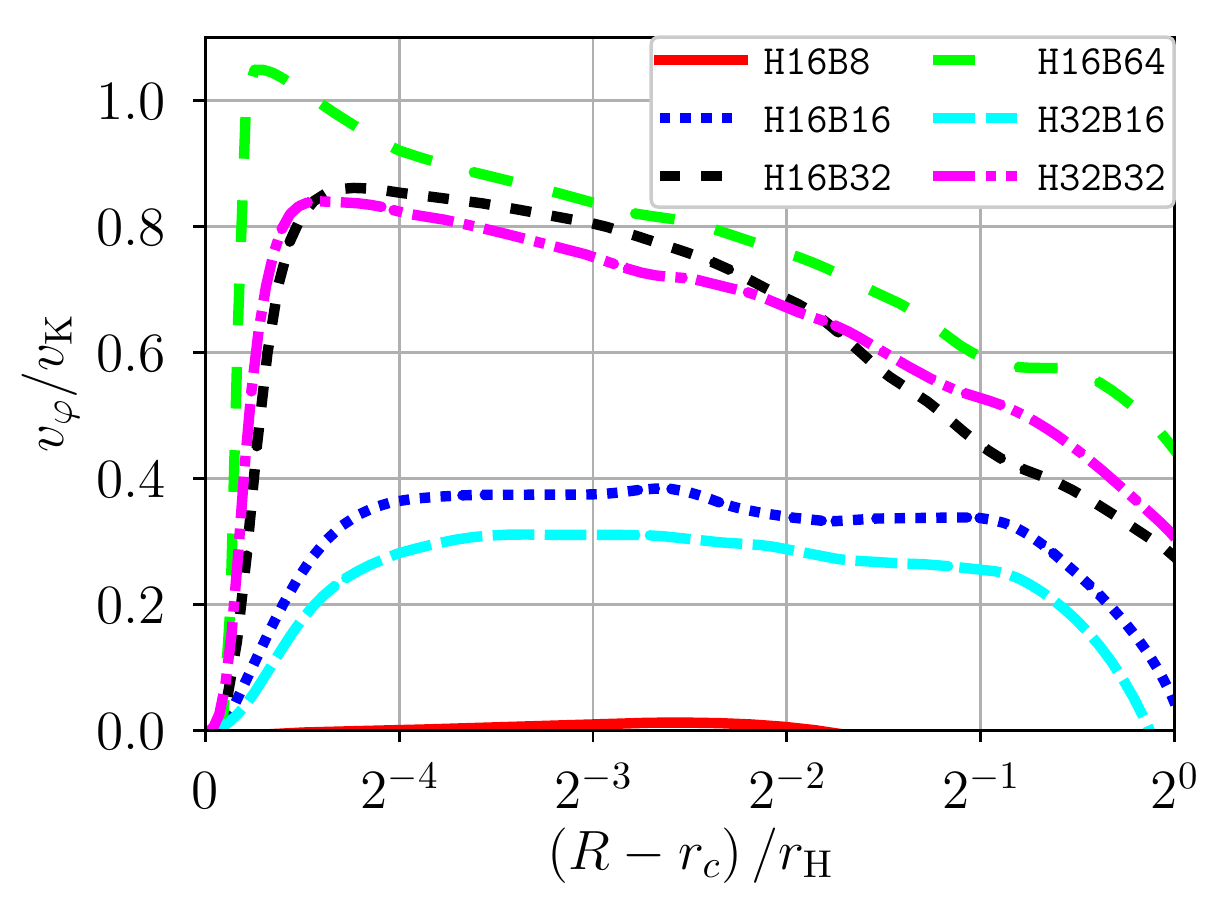}
\caption{Time-averaged radial profiles of the degree of rotational support $v_{\varphi} / \vK$ measured in the equatorial plane of our series of isothermal simulations, each curves corresponding to a different run (see labels). The radial coordinate is measured from the core surface $r_c$ for every run. Note that these radial profiles are primarily functions of $B$ but not of $H$.
\label{fig:i3d_vpp}}
\end{center}
\end{figure}

We now provide a more systematic overview of the role played by the rotational support in determining the structure of the core envelope. \autoref{fig:i3d_vpp} shows the relative measure of rotational support $v_{\varphi}/\vK$ in the equatorial plane of our series of our isothermal simulations; the maximum values of $v_{\varphi}/\vK$ are gathered in \autoref{tab:recap}. This ratio $\approx 1$ per cent is negligible in \texttt{H16B8}, for which the envelope is almost entirely supported by pressure gradients against gravity. The fraction of rotational support grows to approximately $35$ per cent in runs \texttt{H16B16} and \texttt{H32B16}. It reaches $85$ per cent in runs \texttt{H16B32} and \texttt{H32B32}, with the two curves being almost superimposed. As expected from the density contours of \autoref{fig:PLm_rhodv} (d), \texttt{H16B64} achieves full rotational support close to the core. Even in this case, the ratio $v_{\varphi} / \vK$ decreases with radius in the envelope. 

This figure shows that the degree of rotational support in the midplane depends primarily on $B$ and only very weakly on $H$. The definition \eqref{eq:dim-less} then implies that at a given orbital frequency ($\Omega$) and for a given disc temperature ($c_s$), the degree of rotational support of the entire envelope is a function of $m_c/r_c$. One might expect that for $r \gg r_c$ the envelope structure should not depend on the core size anymore as long as $r_c$ is smaller than any other characteristic scales of the problem, such as $r_{\rm B}$ and $h$. In the two-dimensional simulations of Paper I, we examined the dependence of $v_{\varphi}/\vK$ on the size of the core and demonstrated that it should indeed vanish in the limit of $H\gg 1$ (i.e., $r_c\ll h$). In this limit it is natural to expect $v_{\varphi} / \vK$ to depend only on the core \emph{mass}, in which case it would be sensitive to $B/H=m_c/m_{\rm th}$ and not on $B$ alone. However, runs \texttt{H16B8} and \texttt{H32B16} with the same $B/H=0.5$, as well as \texttt{H16B16} and \texttt{H32B32} with $B/H=1$, have very different $v_{\varphi} / \vK$ profiles in \autoref{fig:i3d_vpp}. The core mass is therefore not the only parameter controlling rotational support in these simulations.

On the other hand, we found in Paper I that the convergence rate of $v_{\varphi}/\vK$ with the core size is rather slow. In practice, none of the simulations presented in Paper I or in the current paper effectively reaches this regime with $H\le 32$. \autoref{fig:i3d_vpp} demonstrates that $v_{\varphi} / \vK$ depends on both $m_c$ and $r_c$ (through $B\equiv G m_c / c_s^2 r_c$ mainly), so we must conclude that the cores considered in this study are still large enough to have an effect on their envelope properties. This also implies that the rotational support of the atmospheres of super-Earths discovered by \emph{Kepler} could be noticeably affected by the finite size of the core. 

%%%%%%%%%%%%%%%%%%%%%%%%%%%%%%%%%%%%%%%%%%%%%%%%%%

\subsection{Mass accretion}
\label{sect:mdot}

%%%%%%%%%%%%%%%%%%%%%%%%%%%%%%%%%%%%%%%%%%%%%%%%%%

Here we systematically examine how the mass accretion rate onto the core changes between the different simulation runs. Mass accretion, i.e. the permanent trapping of gas from the disc in the vicinity of the core, is regulated by the energy dissipation of the incoming flow near the core, allowing it to become gravitationally bound. The mass of an isothermal envelope is limited from above only by its hydrostatic value, for which rotational support vanishes. In practice, the steady state mass of the envelope is never reached in our simulations because of their relatively short integration time. 

\begin{figure}%[H]
\begin{center}
\includegraphics[width=1.0\columnwidth]{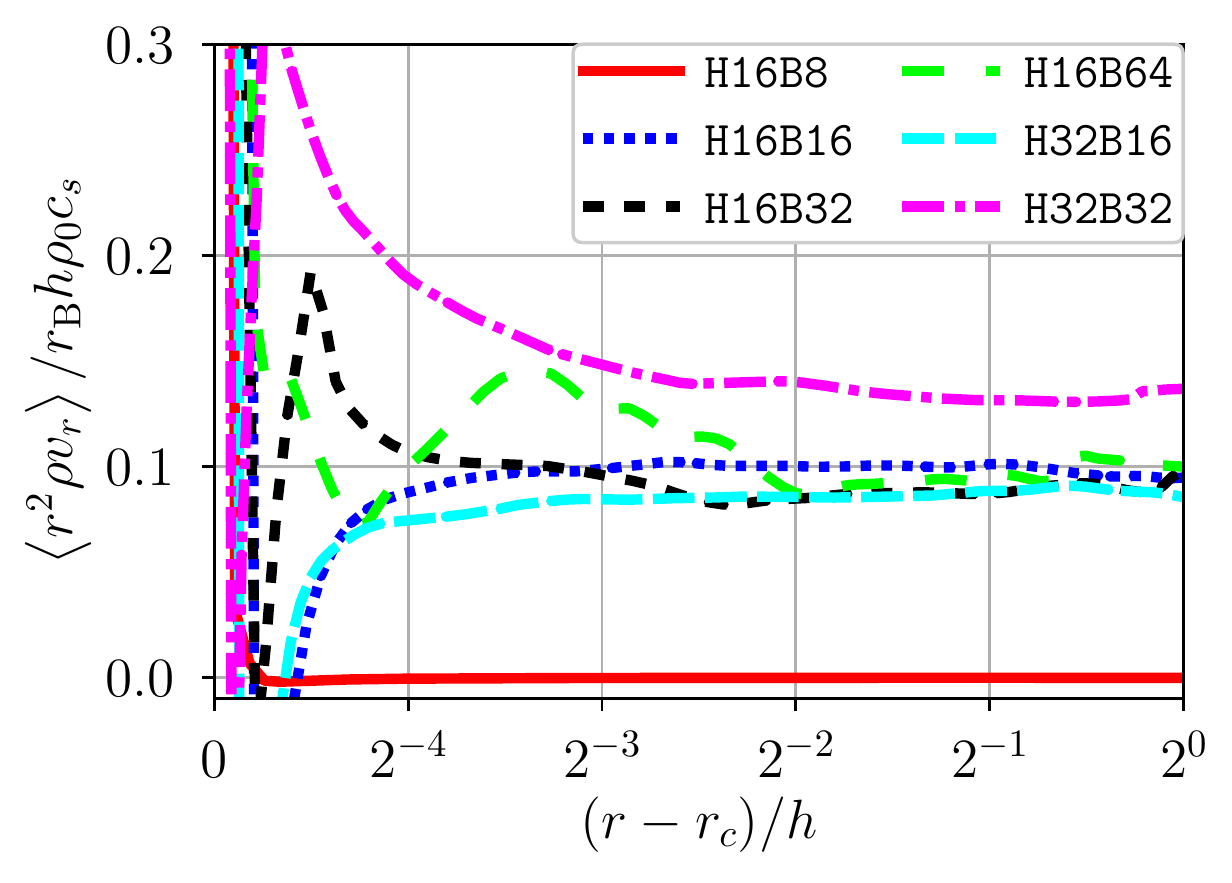}
\caption{Time- and shell-averaged radial profiles of radial mass flux (mass accretion rate). Every run except \texttt{H16B8} has a non-zero and well-defined accretion rate through the envelope, which is roughly the same for all runs when expressed in units of $\rho_0\rB h c_s$. 
\label{fig:i3d_mFr}}
\end{center}
\end{figure}

We display the mass accretion rates measured in our isothermal simulations in \autoref{fig:i3d_mFr}. These accretion rates are calculated by integrating the mass flux over spherical shells of different radii centered on the core; as such they account for both the inflow at the poles as well as the outflow in the equatorial regions. The accretion rates measured this way decrease by less than $0.5$ per cent over the duration of each simulation.

Among the runs presented in this paper, only \texttt{H16B8} has a negligible accretion rate. Incidentally, it is also the only run where the polar inflows do not become supersonic (see \autoref{tab:recap} and \autoref{fig:PLm_rhodv}). This supports the idea that the polar shocks are the primary cause of energy dissipation and mass trapping in the envelope, as opposed to torques in the equatorial plane (cf. Paper I). The polar inflows of run \texttt{H16B8} are channelled toward the midplane with no physical dissipation that would allow the incoming gas to become bound to the core. As a result, the mass accretion rate is compatible with zero in this simulation. An analogous (non-accreting) situation was presented by \cite{ormel2} with $B=10$ and $H\gg B$. 

Moving on to runs with supersonic gas inflows, we find that during the initial stage of envelope accumulation (captured in this study) all such simulations maintain a universal accretion rate 
\ba  
\dot M \approx 0.1\rho_0 \rB h c_s \approx 0.1\Omega^{-1}\rho_0 a^3 \frac{m_c}{m_\star},
\label{eq:dotM}
\ea
irrespective of the parameters $B$ or $H$ as long as $B\gtrsim 16$, see \autoref{fig:i3d_mFr}. 
Equation \eqref{eq:dotM} shows that in runs with supersonic inflows, the accretion rate $\dot M$ scales linearly with the mass of the core and is independent of the thermodynamic conditions in the disc ($\dot M$ is essentially independent of $c_s$). This is different from the standard Bondi accretion rate, for which $\dot M\propto m_c^2 c_s^{-3}$ \citep{bondi52}, because in our case the background state away from the core (the sheared flow of the disc) is different from the static and homogeneous background of \citet{bondi52}. 

To better understand our $\dot M$ behavior we use the results of \citet{Krumholz2005}, who studied Bondi-like accretion in a 3D isothermal flow with non-zero vorticity --- a setup very similar to ours except for the lack of density stratification in the vertical direction, and the absence of the central star (which introduces a Hill sphere and horseshoe orbits into the problem). By demanding the streamline deflection in the vicinity of the accretor to be strong enough to result in shocks and binding of the flow, they arrive at an estimate $\dot M=C\rho_0 \rB h c_s$ in the limit of the flow vorticity $\omega\gtrsim c_s/r_\mathrm{B}$, see their equations (28)-(29). In our case $\omega \sim \Omega$, so this limit corresponds to $r_\mathrm{B}\gtrsim h$, i.e., close to the regime in which the scaling \eqref{eq:dotM} was obtained. The coefficient $C$ in \citet{Krumholz2005} involves additional logarithmic dependence on $\omega r_\mathrm{B}/c_s\gtrsim 1$ and is generally larger than the proportionality constant in \eqref{eq:dotM}. We attribute this difference to a somewhat different setup: unlike \citet{Krumholz2005}, in our case the flow can reach the core only within a limited azimuthal range due to the barrier imposed by the tidal potential. 

If the $\dot M\propto m_c$ behavior given by \eqref{eq:dotM} persists on longer timescales, and if the core is massive enough for its hydrostatic envelope mass to exceed $m_c$, then an exponential growth of the envelope may be possible when it becomes self-gravitating. However, this conclusion applies only to the (rather unrealistic) case of a purely isothermal envelope for which the behavior \eqref{eq:dotM} was established. Moreover, complications such as gap formation are likely to limit mass supply from the disc into the feeding zone of the core on long timescales, slowing down the exponential growth eventually \citep{ginzburgchiang19}. 

%%%%%%%%%%%%%%%%%%%%%%%%%%%%%%%%%%%%%%%%%%%%%%%%%%

\subsection{Envelope recycling}
\label{sec:recycling}

%%%%%%%%%%%%%%%%%%%%%%%%%%%%%%%%%%%%%%%%%%%%%%%%%%

\begin{figure}%[H]
\begin{center}
\includegraphics[width=1.0\columnwidth]{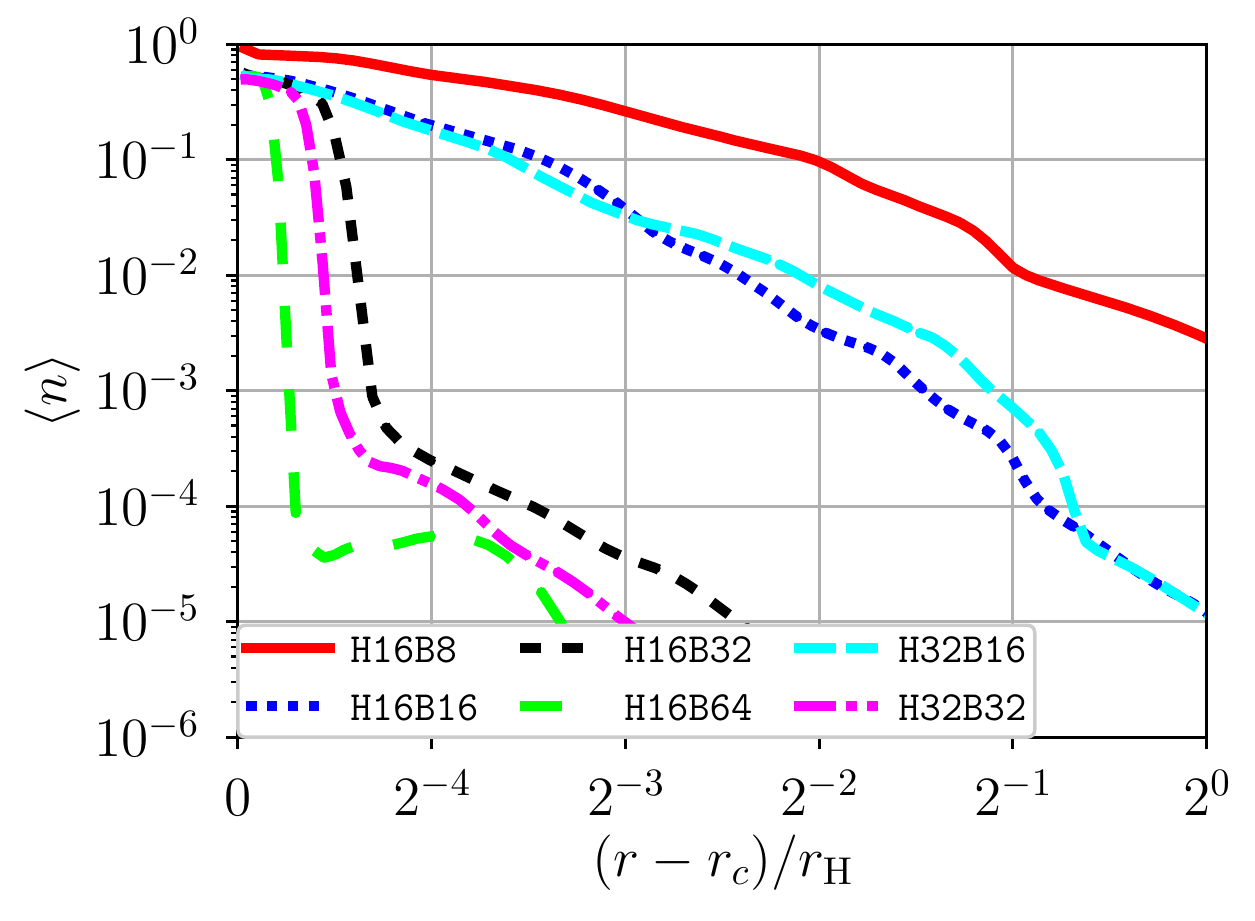}
\caption{Shell-averaged tracer concentration ten orbits after its injection in the Bondi sphere of our isothermal simulations. Different runs are labeled on the panel. As in Figure \ref{fig:i3d_vpp}, the radial profiles of the curves depend primarily on $B$ and only weakly on $H$.
\label{fig:i3d_tr1}}
\end{center}
\end{figure}

We also look at the evolution of the recycling characteristics of the flow within the envelope as a function of $m_c$ and $r_c$. As in \autoref{sec:envrecycle}, we let every simulation settle to a quasi-steady state over ten orbits, and then we inject a passive tracer with constant concentration $n=1$ inside the Bondi sphere. We let the flow evolve for ten additional orbits and draw the resulting concentration profiles on \autoref{fig:i3d_tr1}. Larger final concentrations can be interpreted as longer recycling timescales. 

\autoref{fig:i3d_tr1} shows a pattern of $\langle n(r)\rangle$ very similar to that in \autoref{fig:i3d_vpp}. After ten orbits, run \texttt{H16B8} has an average tracer concentration larger than $10$ per cent inside $r/\rH < 1/4$, corresponding to the longest recycling timescale. The average concentration profiles of runs \texttt{H16B16} and \texttt{H32B16} are nearly superimposed and decrease smoothly with radius, down to $10^{-5}-10^{-4}$ at large radii $r\sim h$. In the remaining runs, the tracer concentration drops by a factor $10^3-10^4$ outside the dense shell surrounding the core, with runs \texttt{H16B32} and \texttt{H32B32} having very similar $\langle n(r)\rangle$ profiles. The polar shocks seem to delimit regions of low (inside) and high (outside) recycling efficiency. The absence of a concentration plateau with $\langle n(r)\rangle=1$ at small radii means that in these runs recycling operates down to the core surface. This progression clearly demonstrates that the profiles of $\langle n(r)\rangle$ depend primarily on the value of $B$ and only weakly on $H$ (as in \autoref{fig:i3d_vpp}). Thus, the finite size of the core, and not only its mass, has an important effect on the recycling properties of the flow. 

We mentioned in \autoref{sec:envrecycle} that the e-folding decay timescale of the cumulative tracer mass may not be well suited to characterize the recycling efficiency of the flow. It is worth looking for additional diagnostics based on the time-averaged flow variables alone \citep[e.g., via streamlines integration,][]{ormel2}. Here we construct one-dimensional recycling diagnostics similar to those of \cite{kurokawa18} in that they do not rely on a tracer fluid. Let $\tilde{X} \equiv X - \brac{X}_{\mathcal{S}}$ be the deviation of a quantity $X$ relative to its shell average at a given radius, and $\| X \|^2 \equiv \oint_{\mathcal{S}} X^2 \,\dd S / 4\pi r^2$ be the squared norm of $X$ on the sphere. Integrating \eqref{eqn:dttr1} over the volume $\mathcal{V}(r)$ of a sphere $\mathcal{S}(r)$ of radius $r$ centered on the core, the enclosed tracer mass evolves according to
\begin{align}
  \frac{\partial}{\partial t} \int_{\mathcal{V}(r)} \!\!\!\!\!\!\! n \rho \,\dd V &= - \oint_{\mathcal{S}(r)} \!\!\!\!\!\!\! n \rho v_r \, \dd S \nonumber\\
&= - 4\pi r^2 \bigl( \brac{n}\cdot\brac{\rho v_r} + \tilde{c}\cdot\|\tilde{n}\|\cdot\wrac{\widetilde{\rho v_r}}\bigr). 
\label{eqn:tracermassphere}
\end{align}
The first product in parenthesis involves the net mass flux across the sphere $\brac{\rho v_r}$. In the second term, $\|\tilde{n}\|$ characterizes the amplitude of variation of the tracer concentration at the surface of the sphere, while $\widetilde{\rho v_r}$ is the dispersion of the radial mass flux on the sphere relative to its shell average \citep[cf. Figure 9 of][]{fung15}. The correlation between $\tilde{n}$ and $\widetilde{\rho v_r}$ on the sphere is measured by $-1\leq\tilde{c}\leq+1$. This correlation remains positive at every radius, but it evolves in time with the tracer distribution in space. 

We define recycling as the replacement of the mass inside a given volume by velocity streams passing through this volume. Recycling conserves the total mass inside the volume and can take place in stationary flows ($\brac{\rho v_r}=0$). For these reasons, we discard the first product in the parenthesis of \eqref{eqn:tracermassphere}. The tracer concentration allows us to track fluid motions, but the recycling properties of the flow should be accessible from the velocity field alone. The only term independent of the tracer distribution is $\wrac{\widetilde{\rho v_r}}$; this term measures the circulation of mass in and out of the sphere, including  both the polar inflows and equatorial outflows. 

\begin{figure}%[H]
\begin{center}
\includegraphics[width=1.0\columnwidth]{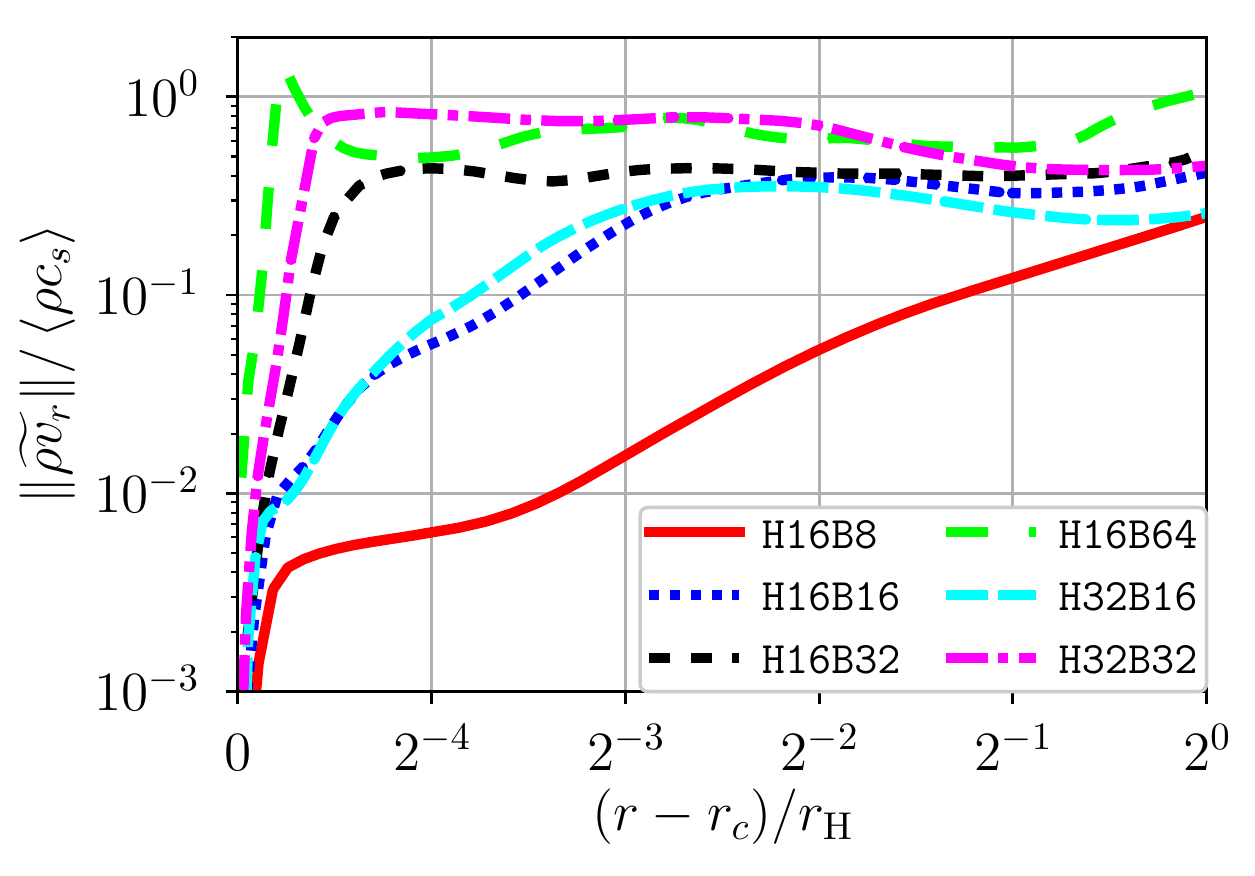}
\caption{Time and shell-averaged absolute radial mass flux deviation $\wrac{\widetilde{\rho v_r}}$ in our series of isothermal simulations, normalised by the shell-averaged $\brac{\rho}c_s$. Note the same hierarchy as in \autoref{fig:i3d_tr1}: $\wrac{\widetilde{\rho v_r}}$ curves depend primarily on $B$ and only weakly on $H$. 
\label{fig:i3d_dpr}}
\end{center}
\end{figure}

\autoref{fig:i3d_dpr} shows the radial profiles of the mean mass flux dispersion $\wrac{\widetilde{\rho v_r}}$ in our simulations. The ordering of these curves is the inverse of \autoref{fig:i3d_tr1}: \texttt{H16B8} is much lower than the other curves, \texttt{H16B16} and \texttt{H32B16} are superimposed, and the three remaining curves feature a strong mass flux dropping only near the core surface. This pattern of dependence primarily on $B$ and not $H$ is again reminiscent of Figures \ref{fig:i3d_vpp} \& \ref{fig:i3d_tr1}. Increasing $B$ enhances the mean radial mass flux dispersion deeper in the envelope, down to the edge of the tracer shells visible in \autoref{fig:i3d_tr1}. For the most massive cores, the mass flux $\wrac{\widetilde{\rho v_r}}$ saturates at a sonic accretion/decretion value $\sim \brac{\rho}_\mathcal{S} c_s$, where $\brac{\rho}_\mathcal{S}$ is the shell-averaged density. The correspondence between \autoref{fig:i3d_tr1} and \autoref{fig:i3d_dpr}, in terms of ordering and transition radii, supports that $\wrac{\widetilde{\rho v_r}}$ essentially captures the recycling efficiency of the flow. 

Assuming that the entire envelope is recycled by the mass flux across its boundary, a recycling timescale can be defined as\footnote{The prefactor $2$ makes this definition exact for a velocity flux tube having a constant mass flux across its section.}
\begin{equation} 
\label{eqn:deftau}
\tau (r) \equiv 2 \frac{\int_{\mathcal{V}(r)} \rho \,\dd V}{\oint_{\mathcal{S}(r)} \wrac{\widetilde{\rho v_r}} \dd S} = 2 \frac{(4/3)\pi r^3 \brac{\rho}_{\mathcal{V}}}{4\pi r^2 \wrac{\widetilde{\rho v_r}}} \gtrsim \frac{\brac{\rho}_{\mathcal{V}}}{\brac{\rho}_{\mathcal{S}}} \frac{r}{c_s},
\end{equation}
where we use the norm $\wrac{\widetilde{\rho v_r}}$ (and its aforementioned saturation at $\brac{\rho}_\mathcal{S} c_s$) instead of the absolute value $\left\vert \widetilde{\rho v_r} \right\vert$ to make a link with \eqref{eqn:tracermassphere} and \autoref{fig:i3d_dpr}. 

We found that the recycling timescale $\tau(r)$ determined this way varies from $30$ to $200$ orbits inside $r<\rH$ in run \texttt{H16B8}, in relatively good agreement with the tracer dispersion timescale determined from \autoref{fig:i3d_tr1}. However, in the fiducial run \texttt{H16B16} we find $\tau>300$ orbits for $r<\rH$. This is more than ten times longer than the tracer dispersion timescale (see \autoref{fig:i3d_tr1}), since in every simulation run except \texttt{H16B8}, the volume-averaged concentration has decreased by $\gtrsim 50$ per cent over ten orbits. This discrepancy results from the inefficient recycling of the shocked shell surrounding the core, whence the numerator of \eqref{eqn:deftau} is largely over-estimated. The volume actually recycled should not include the quasi-static, tracer-rich shells on top of the core. The quantitative agreement between $\tau(r)$ and the tracer dispersion timescale should therefore be limited to low-mass cores, for which the entire envelope is recycled by the mass flux crossing its boundary. 

%%%%%%%%%%%%%%%%%%%%%%%%%%%%%%%%%%%%%%%%%%%%%%%%%%
%%%%%%%%%%%%%%%%%%%%%%%%%%%%%%%%%%%%%%%%%%%%%%%%%%

\section{Discussion} 
\label{sec:discussion}

%%%%%%%%%%%%%%%%%%%%%%%%%%%%%%%%%%%%%%%%%%%%%%%%%%

\subsection{Comparison of 2D and 3D simulations}

In Paper I, we studied the characteristics of the envelopes of embedded planetary cores in a 2D framework, which has also been done before by \cite{miki82}, \cite{kley99}, and \cite{ormel1}. Based on our current results, we now provide a systematic comparison between the 2D and 3D simulation outcomes. We discuss different envelope characteristics, highlighting both the similarities and differences of the 2D and 3D setting, and referring to existing literature. 

The easiest comparison one can make between 2D and 3D setups is in the equatorial plane of the flow. In 3D, the equatorial density distribution has a morphology similar to 2D, although with weaker shocks \citep{bate03}. The equatorial velocity field can still be partitioned into distinct regions --- horseshoe, bulk of the disc, and the envelope itself \citep{fung15}. 

For a given core size and mass, 3D envelopes are more pressure-supported in the equatorial plane than their 2D equivalent. For example, run \texttt{H16B8} has maximum $v_{\varphi} / \vK\approx 70$ per cent in 2D, while in 3D we find maximum $v_{\varphi} / \vK\approx 38$ per cent. This is due to vertical motions bringing mass from the polar cones without accumulating much vorticity \citep{ormel2}. Indeed, azimuthal averaging around the core reveals large vertical motions (see \autoref{fig:PLm_rhodv}) as seen in global simulations of embedded cores \citep{kley01,bate03}. The mean flow is oriented toward the core at high latitudes and radially \emph{outward} in the midplane. In 2D vertical motions are naturally absent, and the flow in the equatorial plane around the core can flow \emph{inward} under gravitational torques, opposite to the 3D isothermal case. 

In 2D mass accretion onto the core must necessarily be accompanied by the loss of angular momentum of the fluid orbiting the core; otherwise, the ``centrifugal barrier'' \citep{ormel1} would not allow accretion to happen. As shown in Paper I, the angular momentum transport is typically effected by the spiral shocks emerging in the circumplanetary disc, although other possibilities exist as well  \citep[e.g., by viscous torques,][]{kley99,lubow99}. 

In our 3D runs the situation is very different. There we find that polar material falls nearly freely towards the core. Depending on the mass of the core, these vertical motions can reach supersonic velocities before shocking on the densest parts of the envelope \citep{ayliffe09,szulagyi17}. Mass accretion is allowed by the dissipation at the standing isothermal shocks in the polar regions \citep{tanigawa12}, which allow the incoming mass to become gravitationally bound to the core. In this way the accreted material thus bypasses the centrifugal barrier, or the ``bottleneck'' of a circumplanetary disc \citep{rivier12}. 

For these reasons, the mass accretion rate $\dot M$ that we find in the 3D case ends up being quite different from its 2D analog. Indeed, we can rewrite \eqref{eq:dotM} as $\dot M\sim \Omega\Sigma a^2(m_c/m_\star)(a/h)$, where $\Sigma\sim \rho_0 h$ is the disc surface density. This form highlights the difference with the behavior suggested in \citet{Tanigawa2016} and \citet{Lee2019} who, based on the 2D isothermal simulations of \citet{Tanigawa2002}, proposed the following scaling: $\dot M\sim \Omega\Sigma a^2(m_c/m_\star)^{4/3}(a/h)^2$. Such a difference with the $\dot M$ behavior found in our 3D isothermal simulations (and interpreted using the analysis of \citealt{Krumholz2005}) may have affected the conclusions of these studies. 

We also find a correlation between the strength of shocks and turbulent variability, mostly excited near the polar axis. Although the time-averaged flow features a smooth sonic surface, the shock surface is prone to variability on short space and time scales. Our 3D isothermal simulations confirm the presence of significant turbulent variability on top of a secular evolution of the envelope \cite{ormel2,nelsonruffert13}.

All of our 3D simulations feature prograde inner envelopes, in agreement with previous numerical studies \citep[e.g.,][]{bate03,wang14}. In 2D setup this property naturally results from the conservation of vortencity \citep[see][and Paper I]{miki82}. However, as we show next, strict vortensity conservation is apparently not necessary to allow formation of the prograde envelopes in a 3D setup. 

%%%%%%%%%%%%%%%%%%%%%%%%%%%%%%%%%%%%%%%%%%%%%%%%%%

\subsection{Vortensity conservation}
\label{sect:vort}

Let $\bm{\omega} \equiv \nabla \times \mathbf{v}$ be the flow vorticity; the potential vorticity, alias vortensity, is defined in a co-rotating frame as $\bm{\varpi} \equiv \left(\bm{\omega}+2\mathbf{\Omega}\right)/ \rho$. The background vortensity in the midplane of a Keplerian shear flow (\ref{eq:shear_flow}) is $\varpi_0 \equiv \Omega/(2\rho_0)$. 

Vortensity is closely related to rotational support in the radial momentum balance \citep[see][and Paper I]{ormel1}. Its evolution in barotropic flows obeys 
\begin{equation} 
\label{eqn:dtvarpi}
\frac{\dd \bm{\varpi}}{\dd t} \equiv \partial_t\bm{\varpi} + (\mathbf{v}\cdot\nabla ) \bm{\varpi} = (\bm{\varpi}\cdot\nabla ) \mathbf{v}.
\end{equation}

In the inviscid 2D case, the vertical component of vortensity $\varpi_z$ is conserved in the Lagrangian sense as the right hand side of equation (\ref{eqn:dtvarpi}) vanishes identically. This conservation of $\varpi_z$ is violated when shocks are present in the flow. In Paper I we demonstrated vortensity production at the spiral shocks launched by the core gravity in a 2D disc. The vortensity perturbation $\delta\varpi=\varpi_z-\varpi_0$ can be positive or negative depending on the geometry of the shock front. However, away from the shocks $\varpi_z$ is strictly conserved in the 2D simulations of Paper I.

\begin{figure}%[H]
\begin{center}
\includegraphics[width=1.0\columnwidth]{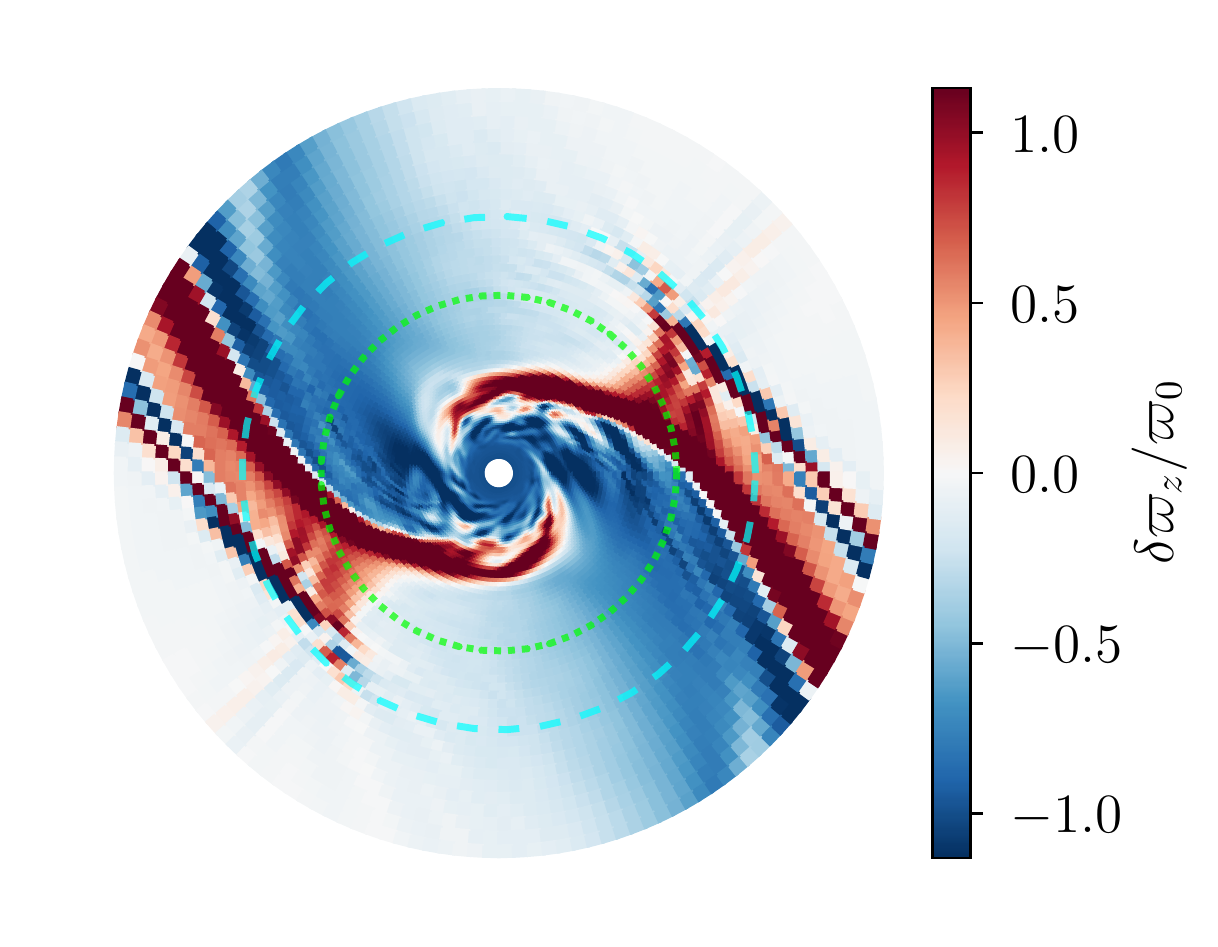}
\caption{Time-averaged deviations $\delta\varpi_z$ of the vertical component of the potential vorticity $\varpi_z$ relative to its background midplane value $\varpi_0=\Omega/(2\rho_0)$ in the equatorial plane of the fiducial run \texttt{H16B16}. The dashed cyan and dotted green circles mark the pressure scale $h$ and the Hill radius $\rH$, respectively. Note that vorticity deviations change sign inside the domain.
\label{fig:i3dp16b16_EQt_wpz}}
\end{center}
\end{figure}

The situation is different in 3D. \autoref{fig:i3dp16b16_EQt_wpz} shows the relative deviations of the vertical vortensity $\delta \varpi_z/\varpi_0 \equiv (\varpi_z-\varpi_0)/\varpi_0$ in the equatorial plane of our fiducial run \texttt{H16B16} ($m_c=m_{\rm th}$). The vortensity itself remains positive everywhere despite order-unity variations, which explains the prograde rotation of the flow around the core in 3D. However, the vortensity deviation $\delta\varpi_z$ exhibits sign variations in different parts of the domain. 

$\delta\varpi_z$ is negative and roughly constant in the innermost region $r\lesssim 2r_c$; every simulation presented here has such an inner envelope with a reduced vortensity. This feature arises because the polar inflows coming from high latitudes bring mass along the direction of unperturbed vortex tubes. Such mass accretion occurs without substantial accumulation of vorticity, reducing the vortensity in this region and causing $\delta\varpi<0$. Mathematically, the negative $\delta\varpi_z$ arises because the right hand side of \eqref{eqn:dtvarpi} is non-zero in 3D and is dominated by the $\varpi_z (\partial v_z/\partial z) < 0$ term in this part of the flow. 

Further away from the core, but still inside the Hill sphere and extending out of the envelope, positive vortensity deviations appear in a non-axisymmetric spiral pattern, with $\delta\varpi_z$ alternating sign at a given radius. Note that these vortensity deviations are not superimposed on the spiral density waves of \autoref{fig:i3dp16b16_EQt_tot}. They again emerge due to the non-zero right-hand side of \eqref{eqn:dtvarpi}. Because the (vertical) inflow velocity $\vert v_z\vert$ decreases with the cylindrical radius $R$, the toroidal vorticity $\omega_{\varphi} \simeq -(\partial v_z/\partial R) > 0$ in a wide cone about the $\theta=0$ axis, and the vortensity $\varpi_\varphi>0$ as well. As shown on \autoref{fig:i3dp16b16_EQt_tot}, the flow is not axisymmetric around the core; in particular, the vertical velocity $v_z$ varies with the azimuthal angle $\varphi$. As a result, the toroidal vorticity $\omega_{\varphi}$ is sheared into a vertical one by the azimuthally-varying vertical flow. Mathematically, the right hand side $(\varpi_{\varphi}/R)(\partial v_z/\partial \varphi)$ is non-zero and alternates sign as a function of the angle $\varphi$. 

This vertical vortensity produced in the Hill sphere is then passively advected by the outward midplane flow into the bulk of the disc. This rather non-trivial sequence of processes explains the variations of $\delta\varpi_z$ in our 3D isothermal simulations.

%%%%%%%%%%%%%%%%%%%%%%%%%%%%%%%%%%%%%%%%%%%%%%%%%%

\subsection{Finite core radius effects}
\label{sect:core-rad}

%%%%%%%%%%%%%%%%%%%%%%%%%%%%%%%%%%%%%%%%%%%%%%%%%%

Under the simplifications of this study, the problem admits two free length scales: the pressure scale $h$ and the Bondi radius $\rB$. It is commonly assumed that the properties of the planet-disc interaction are asymptotically independent of the core size in the limit of small cores ($H\equiv h/r_c\gg1$). Physically, this means that the flow properties should be set by the core mass alone. Because of the relation \eqref{eq:mth-rel} one would then expect $B/H=\rB/h$, and not $B$ or $H$ individually, to determine the envelope characteristics.

Following \cite{ormel1}, in Paper I we examined this assumption via 2D simulations and found that the core radius controls the properties of the entire envelope at least for $H\leq 32$. Using a simplified 1D model based on the assumption of vortensity conservation, we then found that rotating envelopes around smaller cores ($H \sim 10^3$, $B/H\sim 10^{-2}$) could still have a non-vanishing sensitivity to the core radius. 

We make the same observation in our 3D isothermal simulations having $H\leq 32$. Simulations with the same ratio of $B/H$ produce different outcomes depending on the size of the core relative to its Bondi radius. In particular, the degree of rotational support $v_{\varphi}/\vK$, which determines whether a circumplanetary disc forms, depends primarily on $B$ and only moderately on $H$, see \autoref{sect:rot_support}. Similarly, the Mach number of the polar inflows depends primarily on $B$ (\autoref{sect:core_mass_var}). By controlling the strength of the polar shocks, the core radius also influences turbulent variability and mixing properties of the flow (\autoref{sec:recycling}). This sensitivity of the envelope characteristics to $r_c$ is analogous to what we find in 2D, even though arguments based on vortensity conservation no longer apply (see \autoref{sect:vort}).

On the other hand, we find that the mass accretion rate onto the core depends only on the mass of the core to a good accuracy, see \autoref{fig:i3d_mFr} and \autoref{eq:dotM}. We emphasize that this scaling applies only in our twenty orbit-long simulations, when the mass of the envelope is much lower than its equivalent hydrostatic value (see \autoref{sect:mdot}). Nevertheless, this dependence supports the idea that mass accretion rates can be reliably measured in inviscid isothermal simulations even when the core radius is not spatially resolved \citep[see Section 5 in][]{machida10}, as long as accretion is mediated by polar shocks.

%%%%%%%%%%%%%%%%%%%%%%%%%%%%%%%%%%%%%%%%%%%%%%%%%%

\subsection{Recycling properties}

%%%%%%%%%%%%%%%%%%%%%%%%%%%%%%%%%%%%%%%%%%%%%%%%%%

The poloidal circulation in 3D allows material from the envelope to be more efficiently replaced by fresh disc material than in 2D. Indeed, the radial distribution of passive tracer that we find in 2D in Paper I typically exhibits a rather extended plateau near the core with unperturbed tracer distribution, $\langle n\rangle=1$. This plateau can be identified with the rotationally-supported region around the core, which forms a closed system. 

In the 3D case, such a concentration plateau, if present, almost always features $\langle n\rangle$ below unity. This means that recycling operates at some level down to the very surface of the core. The overall morphology of the recycling flows connecting the envelope and the disc is, obviously, also different between 2D (where everything is confined to an equatorial plane) and 3D. In the latter case, the midplane is recycled on orbital timescales, whereas recycling takes several tens of orbits in the polar regions. For sufficiently massive cores inducing polar shocks, the post-shock medium is isolated from the rest of the envelope. Turbulent mixing is inefficient in this innermost shell, so the fluid undergoing the most dissipation is also the least efficiently recycled. 

In 2D, the radial distribution of a passive tracer fluid can, to zeroth order, be considered as depending primarily on $m_c/m_{\rm th}$ --- the non-recycled region is quite extended for $m_c\lesssim m_{\rm th}$, but shrinks substantially for $m_c>m_{\rm th}$. In 3D, as mentioned in \autoref{sect:core-rad}, the recycling characteristics of the flow depend not only on the core mass but also on its radius.

As opposed to the non-isothermal simulations of \citet{ormel3}, \citet{lamblega17}, and \citet{kurokawa18}, in our runs envelope recycling operates efficiently down to the core surface in the absence of shocks (run \texttt{H16B8}) and down to the shocks otherwise. The reduced recycling efficiency reported by \cite{kurokawa18} reflects the formation of a circumplanetary disc in the limit of short cooling time (see their Figure 4). In the simulations of \cite{ormel3} and \cite{lamblega17}, an isolated inner envelope appears due to the efficient gas cooling near the core. The resulting contraction should eventually stop when recycling balances cooling in the entropy budget of the envelope \citep{ormel3}.

%%%%%%%%%%%%%%%%%%%%%%%%%%%%%%%%%%%%%%%%%%%%%%%%%%

\subsection{Transition from sub-thermal to super-thermal mass regimes}

%%%%%%%%%%%%%%%%%%%%%%%%%%%%%%%%%%%%%%%%%%%%%%%%%%

Both Paper I and our present study explored the evolution of envelope characteristics as the core mass was gradually varied from the sub-thermal to the super-thermal regime. Previous studies typically focused on each regime separately \citep{ayliffe09a,ormel1,ormel2,szulagyi16} and employed different setups. This complicates the interpretation of the changes that occur as $m_c$ is varied, something that we do naturally here and in Paper I using a single setup for all values of $m_c$. 

A notable exception is the recent study by \citet{Kuwahara}, who similarly explored the divide between sub-thermal and super-thermal mass cores using a single 3D isothermal numerical setup. The focus of their study was on exploring the evolution of the characteristics of the midplane outflow as $m_c$ was varied; in particular, they demonstrated a significant role played by the midplane outflow in preventing the accretion of solids by the core. Our focus is somewhat different as we concentrate on the gas mass accretion rate, envelope recycling properties, and detailed comparison with the 2D simulations.

We find that both in 2D and 3D the transition from low to high mass cores is accompanied by a considerable change in the envelope properties. Rotational support is insignificant for $m_c<m_{\rm th}$, but becomes very important for $m_c>m_{\rm th}$ both in 2D and 3D, with a rotationally-supported disc forming around the cores with $m_c\gtrsim m_{\rm th}$. In 3D the flow becomes supersonic as the core mass reaches $m_{\rm th}$, starting with polar regions. Radial density distribution, mass accretion, turbulent variability, mixing properties of the flow all change as $m_c$ crosses the threshold at $m_{\rm th}$. At the same time, many characteristics of the flow end up depending not only on $m_c/m_{\rm th}$ but also on the core size, at least for (relatively low) values of $H$ used in this work and Paper I, see \autoref{sect:core-rad}.

%%%%%%%%%%%%%%%%%%%%%%%%%%%%%%%%%%%%%%%%%%%%%%%%%%
%%%%%%%%%%%%%%%%%%%%%%%%%%%%%%%%%%%%%%%%%%%%%%%%%%

\section{Summary} 
\label{sec:summary}

%%%%%%%%%%%%%%%%%%%%%%%%%%%%%%%%%%%%%%%%%%%%%%%%%%

We have studied the gaseous envelopes surrounding embedded planetary cores via inviscid, isothermal, 3D hydrodynamic simulations. Our numerical setup was designed to explicitly include the core as a spatially-resolved impermeable boundary. The simulations were evolved over twenty orbits, allowing us to study the quasi-instantaneous state of embedded planetary atmospheres. Using this setup we probed the transition from low to high-mass cores ranging from $0.5~m_{\rm th}$ to $4~m_{\rm th}$, as well as explored the effect of varying the size of the core. Our main conclusions are as follows.
\begin{enumerate}
\item In agreement with other studies we find that a core embedded in a 3D disc drives a meridional circulation pattern in its Hill sphere: the azimuthally-averaged flow moves toward the core at high latitudes and away from the core in the equatorial plane. The equatorial flow circulates with a prograde orientation around the core, despite the lack of strict conservation of the vertical component of the vortencity. 
\item Similar to the 2D case, we find that for the moderate values of $H\leq 32$ explored in this 3D study the finite size of the core, and not just its mass, affects many properties of the flow. The envelope becomes rotationally supported as $B\equiv \rB/r_c$ increases, with only a weak dependence on $H\equiv h/r_c$. Full rotational support is achieved for $\rB/r_c=64$, but the outer envelope is pressure suppported even then. 
\item The polar inflows fall nearly freely towards high-mass cores. The inflow velocity increases with $B$ and becomes supersonic before shocking on the deep envelope for cores with $B\gtrsim 16$. The dissipation associated with these isothermal shocks allows mass accretion in the vertical direction to occur on orbital timescales; the mass accretion rate scales linearly with the core mass but is independent of the core radius or sound speed in the disc; sonic turbulence appears in the same regime, with turbulent Mach numbers of order unity near the shocks. 
\item Different regions of the envelope are recycled on different timescales, ranging from just a couple of orbits in the midplane region to several tens of orbits near the poles. The shocked gas on top of the core remains dynamically bound to the core while the rest of the envelope is recycled on orbital times for $B\geq 32$. 
\end{enumerate}

The main simplification of this study is the isothermal assumption for the entire flow. Choosing a different equation of state appears to affect the circulation of the flow in inviscid simulations where the embedded core is spatially resolved \citep{ormel3}. In particular, \cite{kurokawa18} report that recycling can be largely suppressed in adiabatic flows due to a buoyancy barrier. The inclusion of a more sophisticated treatment of the gas thermodynamics is the natural extension of the current work, which will be presented in a future study. 

\section*{Acknowledgements}

Financial support of this work by the Isaac Newton Trust, Department of Applied Mathematics and Theoretical Physics and STFC through grant ST/P000673/1 is gratefully acknowledged. We thank the referee for making a number of useful comments and suggestions to improve this paper. W.B. thanks Richard Nelson and Pablo Benitez-Llambay for sharing their insight on this topic. 

%%%%%%%%%%%%%%%%%%%%%%%%%%%%%%%%%%%%%%%%%%%%%%%%%%



\bibliographystyle{mnras}
\bibliography{biblio} % if your bibtex file is called example.bib

%%%%%%%%%%%%%%%%%%%%%%%%%%%%%%%%%%%%%%%%%%%%%%%%%%

%%%%%%%%%%%%%%%%% APPENDICES %%%%%%%%%%%%%%%%%%%%%

%\appendix


%%%%%%%%%%%%%%%%%%%%%%%%%%%%%%%%%%%%%%%%%%%%%%%%%%

% Don't change these lines
\bsp	% typesetting comment
\label{lastpage}
\end{document}